\documentclass[ reprint,aps]{revtex4-2}
\usepackage{amssymb}
\usepackage{amsmath}
\usepackage{tabularx,epsfig,color,tabularx,epstopdf}
\usepackage{graphicx}
\usepackage{dcolumn}
\usepackage{bm}

\begin{document}

\title{Time-localized dark modes generated by zero-wavenumber-gain
modulational instability}
\author{Lei Liu$^{1}$}
\author{Wen-Rong Sun$^{2}$}
\email{Corresponding author: sunwenrong@ustb.edu.cn}
\author{Boris A. Malomed$^{3,4}$}
\author{P.G. Kevrekidis$^{5}$}
\affiliation{$^{1}$College of Mathematics and Statistics, Chongqing University, Chongqing, 401331, China\\
$^{2}$School of Mathematics and Physics, University of Science and Technology Beijing, Beijing 100083, China\\
$^{3}$Department of Physical Electronics, School of Electrical Engineering, Faculty of Engineering, and Center for Light-Matter Interaction, Tel Aviv University, P.O.B. 39040, Ramat Aviv, Tel Aviv, Israel\\
$^4$Instituto de Alta Investigaci\'{o}n, Universidad de Tarapac\'{a},
Casilla 7D, Arica, Chile\\
$^5$ Department of Mathematics \& Statistics, University of
Massachusetts, Amherst, MA 01003, USA} 

\begin{abstract}
We report the emergence of a novel type of solitary waves, \textit{viz}.,
\emph{time-localized dark modes }in integrable and non-integrable variants
of the massive Thirring model and in the three-wave resonant-interaction
system, which are models broadly used in plasma physics, nonlinear optics,
and hydrodynamics. They are also interesting as basic models for the
propagation of nonlinear waves in media without intrinsic dispersion. An
essential finding is that the condition for the existence of time-localized
dark modes in these systems, which develop density dips in the course of
their evolution, coincides with the condition for the occurrence of the
zero-wavenumber-gain (ZWG) modulational instability (MI). Systematic
simulations reveal that, whenever the ZWG MI is present, such dark modes are
generically excited from a chaotic background as patches embedded in complex
patterns. 
\end{abstract}

\maketitle


\preprint{APS/123-QED}






\section{ Introduction}

The modulational instability (MI) of a constant-amplitude continuous-wave
(CW) background against long-wavelength perturbations is a fundamental
phenomenon in nonlinear physics~\cite{mi1,mi7,xj1,xj2}. It triggers complex
dynamics in water waves~\cite{mi1,mi7,water}, plasmas~\cite{yy1,yy2,yy3},
electric transmission lines \cite{yy5,Kengne}, nonlinear optics \cite%
{Hasegawa}-\cite{KartSkr}, matter waves \cite{KonSal}-\cite{Ishfaq}, and
other physical media \cite{Kamchatnov,granular}. In particular, MI initiates
the spontaneous production of self-sustained states, such as soliton trains,
breathers, and rogue waves (RWs) \cite%
{science1,mi2,mi3,mi4,mi5,zp1,zp2,zp4,zp5,zp6,zp8,rw18,rw19,rw20,K1}.

Similar to MI, solitons are formed as a result of the interplay between
dispersive and nonlinear effects~\cite{js1}. Universal integrable models,
such as the Korteweg-de Vries and nonlinear Schr\"{o}dinger (NLS) equations
and the Manakov system, give rise to the commonly known exact solutions for
solitons~\cite{js2,js22}. Solutions for traveling solitons can be often
generated by the application of a suitable (Galilean or Lorentz) boost to
quiescent ones. However, conservation laws (first of all, the conservation
of the total norm) imply that the NLS or similar integrable equations do not
admit the existence of time-localized (pulsed) states \cite{zp8, rw18}. It
may seem that the existence of RWs contradicts this statement, as apparent
localization in time $t$ is their basic feature \cite{rw12,rw9}. However,
unlike bright solitons, RWs exist on top of a CW background, and, at fixed $%
t $, RW solutions feature local intensity values below and above the CW
level in a mutually compensating way, which makes them compatible with the
underlying model conservation laws.

In this work, we use two basic integrable systems, \textit{viz}., the
massive Thirring model (MTM) and three-wave resonant-interaction (3WRI)
system, to produce novel waveforms in the form of dark time-localized modes,
which similar to the spatial structure of dark solitons, feature a
time-localized dip in the course of their evolution. An important
observation is that the existence condition for such temporarily-dark
solutions in these systems coincides with the condition of the presence of
the zero-wavenumber-gain (ZWG) MI, i.e., MI with nonzero gain at the zero
wavenumber of modulational perturbations, defined as in~Ref. \cite{LWB2022}.
Moreover, the same systems admit configurations built as multiple sets of
such modes, in compliance with the conservation loss. To the best of our
knowledge, the present work is the first one to show the existence and
origin of time-localized dark and anti-dark modes (the latter meaning states
with a temporarily localized bulge on top of the CW background).

The rest of the paper is organized as follows. Exact time-localized
solutions if the integrable MTM are produced in Section 2. The analytical
investigation of the MI of the flat CW states, with emphasis on the case of
the ZWG MI, is presented in Section 3. Numerical results, which display the
generation of complex patterns, which include local patches of
time-localized modes, by random perturbations initially added to the CW
background, are summarized in Section 4. The other integrable model, whose
exact solutions also demonstrate time-localized modes, \textit{viz}., the
three-wave resonant-interaction system, is briefly considered in Section 5.
The paper is concluded by Section 6.

\section{Time-localized dark modes produced by the MTM}

The MTM system, written in the laboratory coordinates, applies to the
evolution of a self-interacting spinor field in the one-dimensional field
theory \cite{WT1959,AD2015} and constitutes the integrable model which is
most proximal to, but different from, the system governing the propagation
of light in fiber Bragg gratings~\cite{DR1989,Aceves,AHA1997,ASA2015,AD2015}%
. The scaled the form of MTM\ is:
\begin{subequations}
\label{mt1}
\begin{eqnarray}
&&i\partial _{t}u_{1}+i\partial _{x}u_{1}+u_{2}+|u_{2}|^{2}u_{1}=0, \\
&&i\partial _{t}u_{2}-i\partial _{x}u_{2}+u_{1}+|u_{1}|^{2}u_{2}=0.
\end{eqnarray}%
Here $u_{1}$ and $u_{2}$ are slowly varying complex envelopes of
counterpropagating electromagnetic waves (in terms of optics), $t$ and $x$
are the normalized time and spatial coordinate, with the group velocities
and nonlinearity coefficient scaled to be, respectively, $\pm 1$ and $1$.
Note that Eqs.~(\ref{mt1}) can be written in another well-known form in
terms of the light-cone coordinates, $\left( x\pm t\right) /\sqrt{2}$~\cite%
{ASA2015,JBB2022,JB2022}, and can be transformed into the single sine-Gordon
equation, which is integrable too \cite{Coleman}.

General $N$-bright and $N$-dark soliton solutions of the MTM in the
light-cone coordinates can be produced by the Hirota bilinear method~\cite%
{JB2022}. We find that, differently from conventional solitons, dark and
anti-dark soliton solutions of Eqs.~(\ref{mt1}) in the laboratory
coordinates admit a time-localized shape. Note that the MTM does not admit
time-localized bright and dark solitons in the light-cone coordinates, and
bright solitons of Eqs.~(\ref{mt1}) cannot be time-localized either \cite%
{JB2022} .

Fundamental dark- or anti-dark-mode solutions of Eqs.~(\ref{mt1}) are
written as \cite{JB2022}
\end{subequations}
\begin{subequations}
\label{dd1}
\begin{eqnarray}
&&\hspace{-0.5cm}u_{1}=a_{1}e^{i\theta (x,t)}\frac{1+e^{\xi _{1}+\xi
_{1}^{\ast }+i\phi _{1}+\kappa _{1}}}{1+e^{\xi _{1}+\xi _{1}^{\ast }+\kappa
_{1}}}=a_{1}e^{i\theta (x,t)}\times   \notag \\
&&\hspace{0.2cm}\left[ 1+e^{i\phi _{1}}+(e^{i\phi _{1}}-1)\text{tanh}(\xi
_{1}+\xi _{1}^{\ast }+\frac{\kappa _{1}}{2})\right] , \\
&&\hspace{-0.5cm}u_{2}=a_{2}e^{i\theta (x,t)}\frac{1+e^{\xi _{1}+\xi
_{1}^{\ast }+i\phi _{2}+\kappa _{1}}}{1+e^{\xi _{1}+\xi _{1}^{\ast }+\kappa
_{1}^{\ast }}}=a_{2}e^{i\theta (x,t)}\times   \notag \\
&&\hspace{0.2cm}\left[ 1+e^{i\phi _{2}}+(e^{i\phi _{2}}-1)\text{tanh}(\xi
_{1}+\xi _{1}^{\ast }+\frac{\kappa _{1}^{\ast }}{2})\right] ,
\end{eqnarray}%
where
\end{subequations}
\begin{equation}
\theta (x,t)=\frac{1}{2}(1+a_{1}a_{2})\left[ (\frac{a_{2}}{a_{1}}-\frac{a_{1}%
}{a_{2}})x+(\frac{a_{2}}{a_{1}}+\frac{a_{1}}{a_{2}})t\right] ,  \label{theta}
\end{equation}%
\begin{gather*}
e^{\kappa _{1}}=-\frac{ip_{1}^{\ast }}{p_{1}+p_{1}^{\ast }}, \\
e^{i\phi _{1}}=-\frac{p_{1}-i\beta }{p_{1}^{\ast }+i\beta }, \\
e^{i\phi _{2}}=-\frac{p_{1}-i\beta (1+a_{1}a_{2})}{p_{1}^{\ast }+i\beta
(1+a_{1}a_{2})}, \\
\xi _{1}=\frac{\chi _{1}}{2}x+\frac{\chi _{2}}{2}t+\xi ^{(0)},
\end{gather*}%
\begin{equation}
\chi _{j}=\frac{a_{2}}{\beta a_{1}}p_{1}-\left( -1\right) ^{j}\frac{\beta
a_{1}}{a_{2}}(1+a_{1}a_{2})p_{1}^{-1},~j=1,2.  \label{chi}
\end{equation}%
Here $\ast $ stands for the complex conjugate, while $p_{1}$, $\xi ^{(0)}$
and $a_{1}$, $a_{2}$, $\beta $ are complex and real constants, respectively,
which must satisfy the following constraint:
\begin{equation}
|p_{1}-i\beta (1+a_{1}a_{2})|^{2}=\beta ^{2}a_{1}a_{2}(1+a_{1}a_{2}).
\label{dd2}
\end{equation}%
If we separate the real and imaginary parts in the complex parameter, $%
p_{1}\equiv p_{1R}+ip_{1I}$, the component $u_{1}(x,t)$ of the solution
exhibits a temporarily dark-mode shape for $\beta p_{1R}<0$, and an antidark
one in the opposite case, while $u_{2}$ represents a dark-mode shape at $%
\beta (1+a_{1}a_{2})p_{1R}<0$ and an antidark-mode one in the opposite case.

\begin{figure}[tbp]
\centering
\includegraphics[height=85pt,width=110pt]{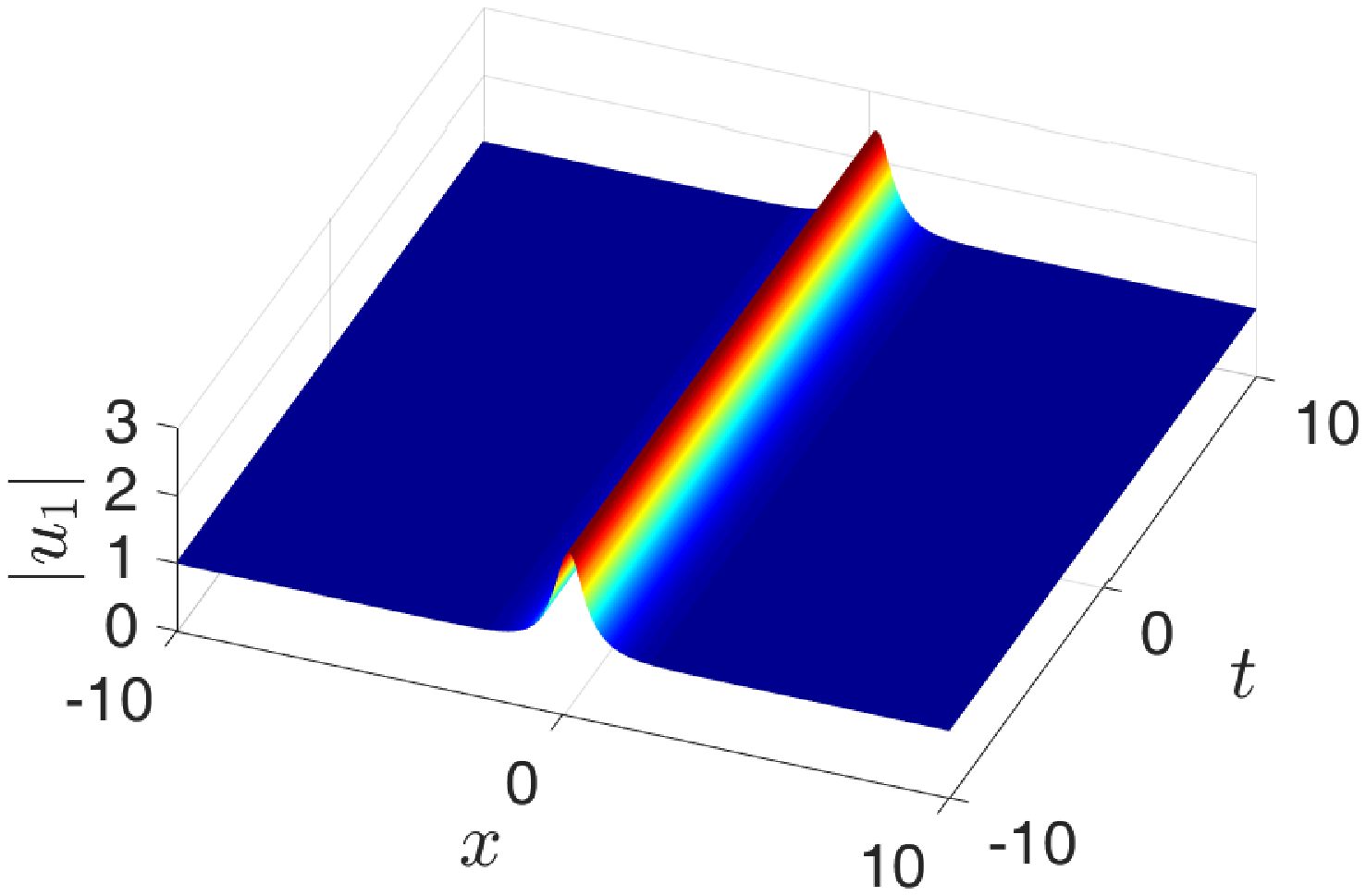}\hspace{0.6cm}%
\includegraphics[height=85pt,width=110pt]{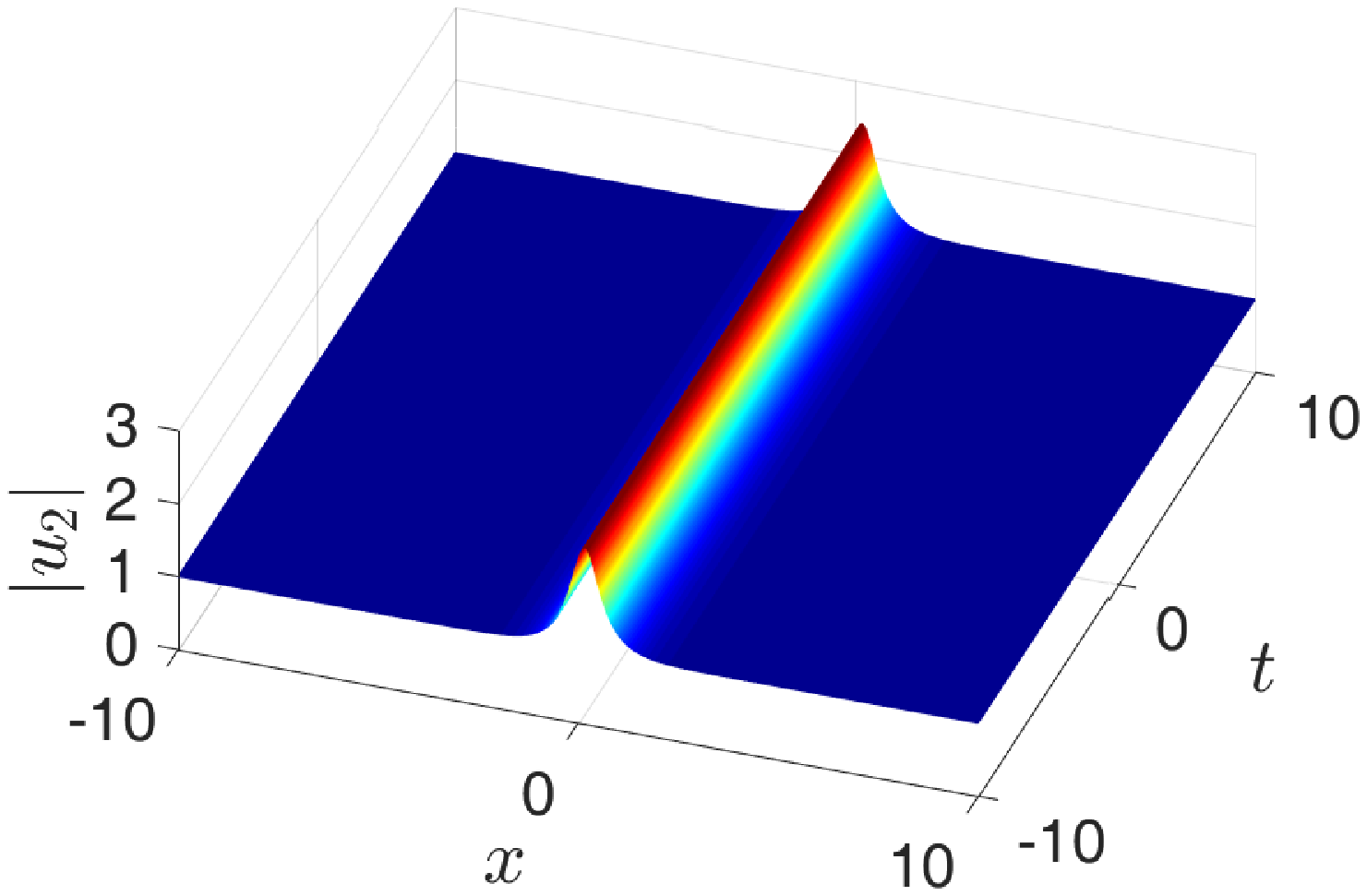} \newline
\vspace{0.2cm}{\hspace{0.9cm}{\footnotesize (a)\hspace{4cm}(b)}}\newline
\vspace{0.2cm} \includegraphics[height=85pt,width=110pt]{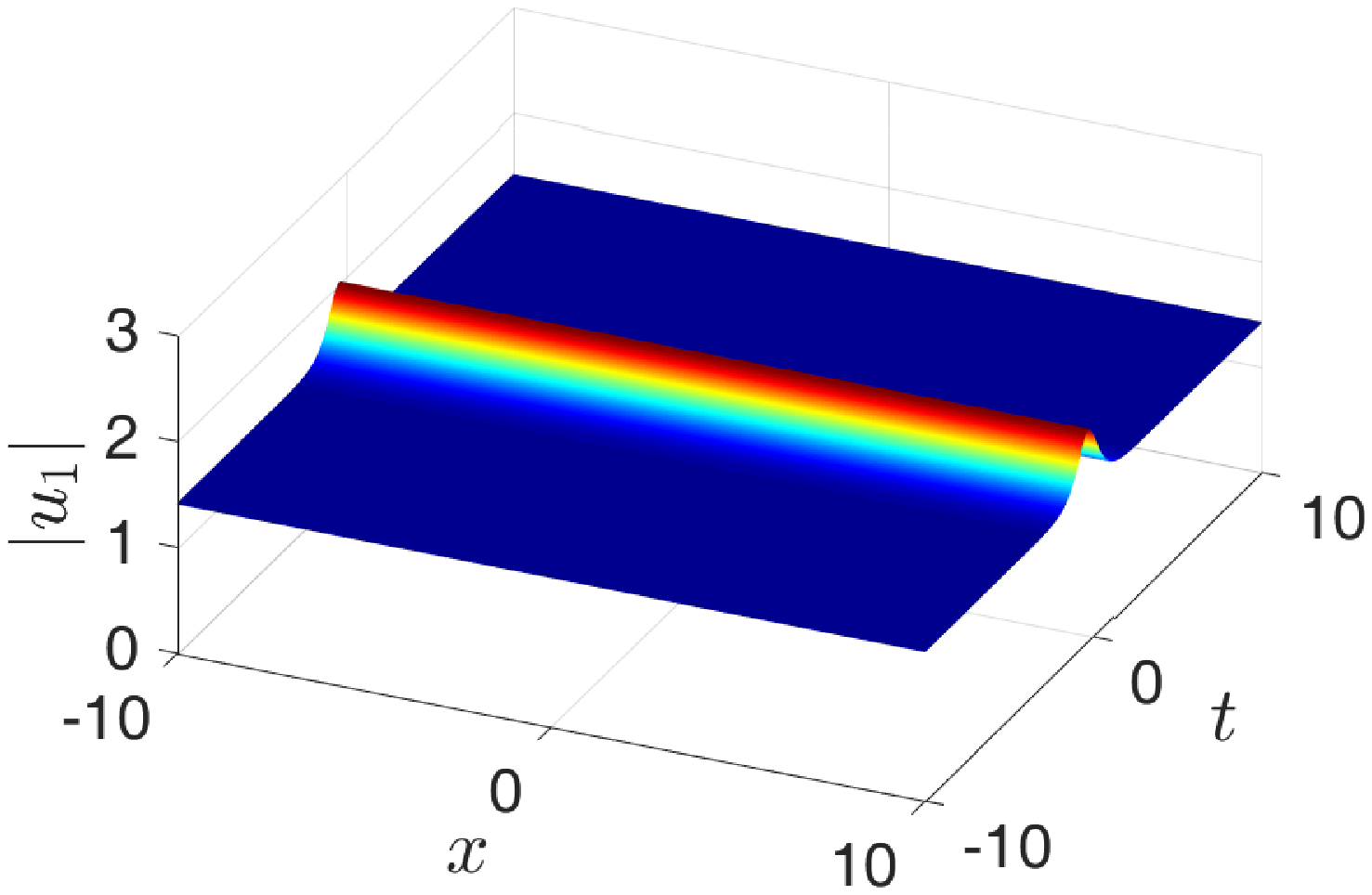}%
\hspace{0.6cm}\includegraphics[height=85pt,width=110pt]{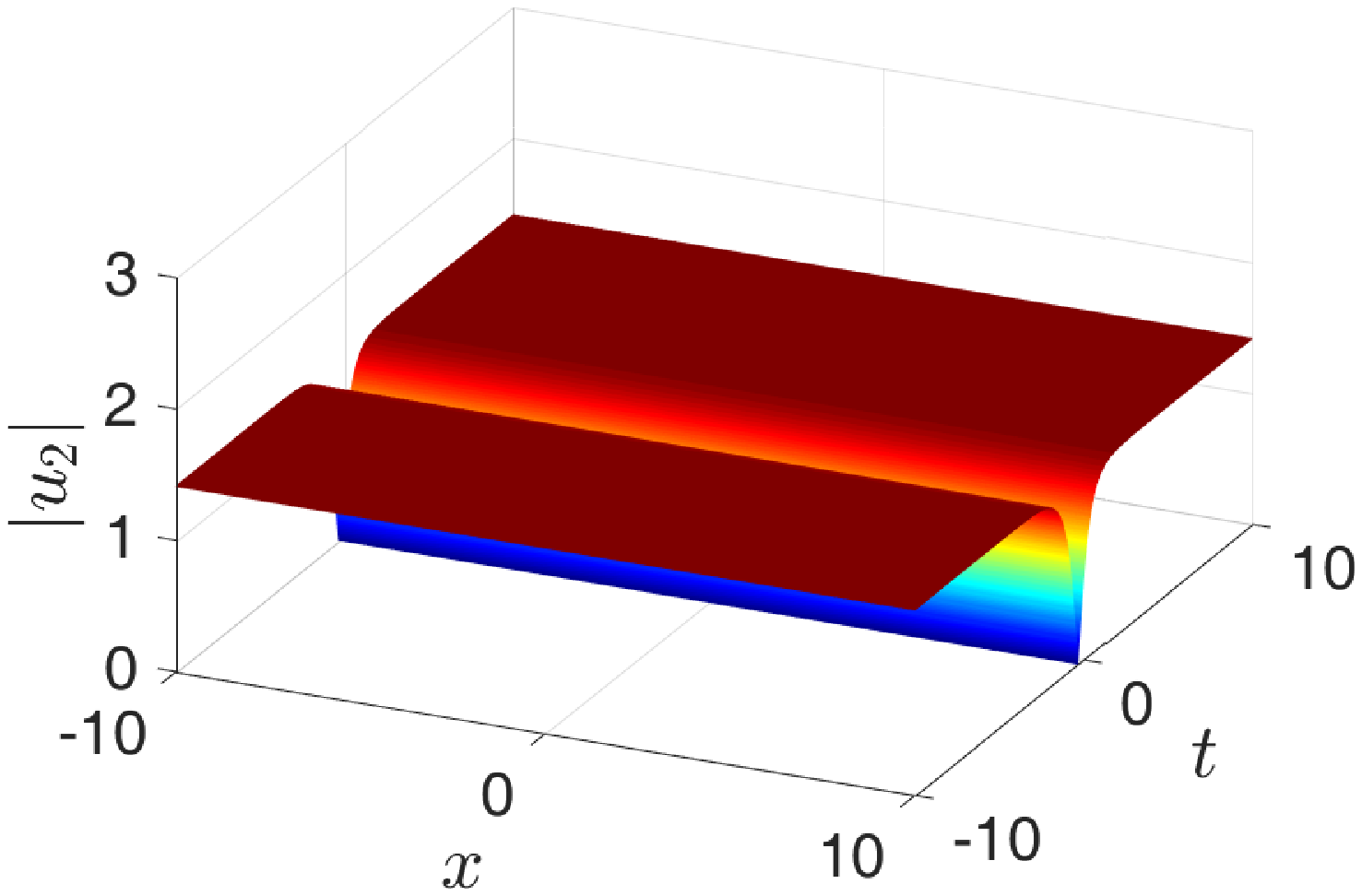}
\newline
\vspace{-0.2cm}{\hspace{0cm}{\footnotesize (c)\hspace{4cm}(d)}}
\caption{Solutions produced by Eqs. (\protect\ref{mt1}) with parameters $%
\protect\beta =1$, $\protect\xi ^{(0)}=0$. (a) and (b): A stationary
two-component spatial anti-dark soliton with $a_{1}=a_{2}=1$ and $p_{1}=1+i$%
. (c) and (d): A time-localized half-anti-dark half-dark solution with $%
a_{1}=-a_{2}=\protect\sqrt{2}$ and $p_{1}=1$.}
\label{fig1}
\end{figure}

The expression (\ref{chi}) for the fully time-localized dark or anti-dark
solutions is
\begin{equation}
\hspace{-0.17cm}\chi _{1}+\chi _{1}^{\ast }=\frac{p_{1}+p_{1}^{\ast }}{\beta
|p_{1}|^{2}a_{1}a_{2}}\left[ \beta
^{2}a_{1}^{2}(1+a_{1}a_{2})+a_{2}^{2}|p_{1}|^{2}\right] =0.
\label{infconstraint1}
\end{equation}%
Combining Eqs.~(\ref{dd2}) and~(\ref{infconstraint1}), we then obtain
\begin{equation}
\beta (1+a_{1}a_{2})[2p_{1I}a_{2}^{2}+\beta (a_{1}^{2}-a_{2}^{2})]=0.
\label{infconstraint3}
\end{equation}%
From Eq.~(\ref{dd2}), we get $\beta (1+a_{1}a_{2})\neq 0$, hence Eq. (\ref%
{infconstraint3}) yields $2p_{1I}a_{2}^{2}+\beta (a_{1}^{2}-a_{2}^{2})=0$,
which further results in
\begin{subequations}
\label{infconstraint4.5}
\begin{eqnarray}
&&p_{1I}=-\frac{\beta (a_{1}^{2}-a_{2}^{2})}{2a_{1}^{2}}, \\
&&p_{1R}=\pm \frac{|\beta a_{1}a_{2}|}{2a_{2}^{2}}\sqrt{-\left( 2+\frac{%
a_{1}^{2}}{a_{2}^{2}}+\frac{a_{2}^{2}}{a_{1}^{2}}+4a_{1}a_{2}\right) }.
\end{eqnarray}%
Because $p_{1R}$ is a nonzero real constant, parameters $a_{1}$ and $a_{2}$
need to satisfy the constraint
\end{subequations}
\begin{equation}
2+\frac{a_{1}^{2}}{a_{2}^{2}}+\frac{a_{2}^{2}}{a_{1}^{2}}+4a_{1}a_{2}<0.
\label{infconstraint4}
\end{equation}%
In other words, inequality~(\ref{infconstraint4}) is the existence condition
for the time-localized dark modes, where $a_{1}$ and $a_{2}$ represent the
background amplitudes of the dark-mode components $u_{1}$ and $u_{2}$,
respectively. On the other hand, the stationary dark-mode solution can be
obtained by setting $\chi _{2}$ to be purely imaginary. Figures~\ref{fig1}%
(a,b) and (c,d) display, severally, examples of stationary states which
feature the spatially-localized anti-dark shape in both components (i.e., it
is a two-component spatial anti-dark soliton), or the anti-dark
temporarily-localized shape in component $u_{1}$, and the temporal dark
shape in $u_{2}$. The former solution is displayed for the sake of the
comparison of the spatial solitons with the time-localized modes.

\section{The linear-stability analysis of CW solutions and the ZWG MI
condition}

Equations~(\ref{mt1}) admit the following CW solutions:
\begin{equation}
u_{l}=a_{l}e^{i[\theta (x,t)+\theta _{0}]},\hspace{0.5cm}l=1,2,
\end{equation}%
where $\theta (x,t)$ is defined as per Eq.~(\ref{theta}), and $\theta _{0}$
is a real phase shift. To study the linear stability of the CW, we add small
complex perturbations $p_{l}(x,t)$ to it, setting
\begin{equation}
\widetilde{u_{l}^{p}}=[a_{l}+p_{l}(x,t)]e^{i[\theta (x,t)+\theta _{0}]},%
\hspace{0.5cm}l=1,2.  \label{pwp}
\end{equation}%
Substituting expressions~(\ref{pwp}) in Eqs.~(\ref{mt1}), we derive
linearized equations for $p_{l}(x,t)$,
\begin{subequations}
\label{2}
\begin{eqnarray}
&&ia_{1}\partial _{t}p_{1}+ia_{1}\partial
_{x}p_{1}-a_{2}p_{1}+a_{1}(1+a_{1}a_{2})p_{2}+a_{1}^{2}a_{2}p_{2}^{\ast }=0,
\notag \\
&& \\
&&ia_{2}\partial _{t}p_{2}-ia_{2}\partial
_{x}p_{2}-a_{1}p_{2}+a_{2}(1+a_{1}a_{2})p_{1}+a_{1}a_{2}^{2}p_{1}^{\ast }=0.
\notag \\
&&
\end{eqnarray}%
Assuming, as is customary, $p_{l}=\eta _{l,1}(t)e^{iQx}+\eta
_{l,2}(t)e^{-iQx}$, where $Q$ is a real perturbation wavenumber, and $\eta
_{l,1}(t)$, $\eta _{l,2}(t)$ are complex amplitudes, Eq. (\ref{2}) leads to
a $4\times 4$ homogeneous linear differential equation in the matrix form
for $\eta =(\eta _{1,1},\eta _{1,2}^{\ast },\eta _{2,1},\eta _{2,2}^{\ast
})^{T}$ as
\end{subequations}
\begin{equation}
\partial _{t}\eta =i\mathbf{M}\eta ,
\end{equation}%
where the matrix elements of $\mathbf{M}$ are $M_{11}=-Q-a_{2}/a_{1}$, $%
M_{22}=-Q+a_{2}/a_{1}$, $M_{33}=Q-a_{1}/a_{2}$, $M_{44}=Q+a_{1}/a_{2}$, $%
M_{41}=M_{32}=-M_{23}=-M_{14}=a_{1}a_{2}$, $%
M_{13}=M_{31}=-M_{24}=-M_{42}=1+a_{1}a_{2}$, $M_{12}=M_{21}=M_{34}=M_{43}=0.$

The stability of solution~(\ref{pwp}) is then determined by eigenvalues of
matrix $\mathbf{M}$, which are roots of the following characteristic
polynomial,
\begin{equation}
\Omega ^{4}+\lambda _{2}\Omega ^{2}+\lambda _{1}\Omega +\lambda _{0}=0,
\label{5}
\end{equation}%
where we define
\begin{gather*}
\lambda _{0}=Q^{2}\left( -\frac{a_{1}^{2}}{a_{2}^{2}}-\frac{a_{2}^{2}}{%
a_{1}^{2}}+4a_{1}a_{2}+Q^{2}+2\right) , \\
\lambda _{1}=2Q\left( \frac{a_{2}^{2}}{a_{1}^{2}}-\frac{a_{1}^{2}}{a_{2}^{2}}%
\right) , \\
\lambda _{2}=-2(1+Q^{2})-4a_{1}a_{2}-\frac{a_{1}^{2}}{a_{2}^{2}}-\frac{%
a_{2}^{2}}{a_{1}^{2}}.
\end{gather*}%
Roots of Eq.~(\ref{5}) ($\Omega _{j},j=1,2,3,4$) are either real ones, or
form complex-conjugate pairs. If all the roots are real, there is no MI. If
frequencies $\Omega _{j}$ include complex-conjugate pairs, MI is represented
by $\text{Im}\left( \Omega \right) <0$. Similar to the setting considered in
Ref. \cite{LWB2022}, MI may be of three different types, \textit{viz}.,
\newline
$\bullet $ Baseband MI: $\text{Im}(\Omega )<0$ at $|Q|>0$ and $\text{Im}%
(\Omega )=0$ at $Q=0$, i.e., the MI band includes arbitrarily small
wavenumbers $Q$ but \emph{not} $Q=0$. \newline
$\bullet $ Passband MI: $\text{Im}(\Omega )<0$ at $|Q|>Q_{\min }>0$ with a
nonzero boundary $Q_{\min }$ of the MI band, which separates it from $Q=0$.%
\newline
$\bullet $ ZWG MI: $\text{Im}(\Omega )<0$ at $|Q|<Q_{\max }$ with $Q_{\max
}>0$, i.e., the MI\ band \emph{includes} zero wavenumber, $Q=0$.

When the MI exists, the boundaries of the ZWG-MI region are defined by
setting $Q=0$ in Eqs.~(\ref{5}). Then, two possible nonzero roots of Eqs.~(%
\ref{5}) are $\pm \sqrt{\Omega _{0}^{2}}$, with
\begin{equation}
\Omega _{0}^{2}=2+\frac{a_{1}^{2}}{a_{2}^{2}}+\frac{a_{2}^{2}}{a_{1}^{2}}%
+4a_{1}a_{2}.  \label{zwg}
\end{equation}%
The ZWG MI takes place at $\Omega _{0}^{2}<0$, otherwise there may exist
only baseband or passband MI regions. We stress that this condition \emph{%
coincides} with the existence condition for the time-localized dark mode,
which is given above by Eq.~(\ref{infconstraint4}). This fact strongly
indicates that the emergence of time-localized modes is intimately
connected with the growth of the modulational perturbation with wavenumber $%
Q=0$.

\begin{figure}[tbp]
\centering
\includegraphics[height=97pt,width=115pt]{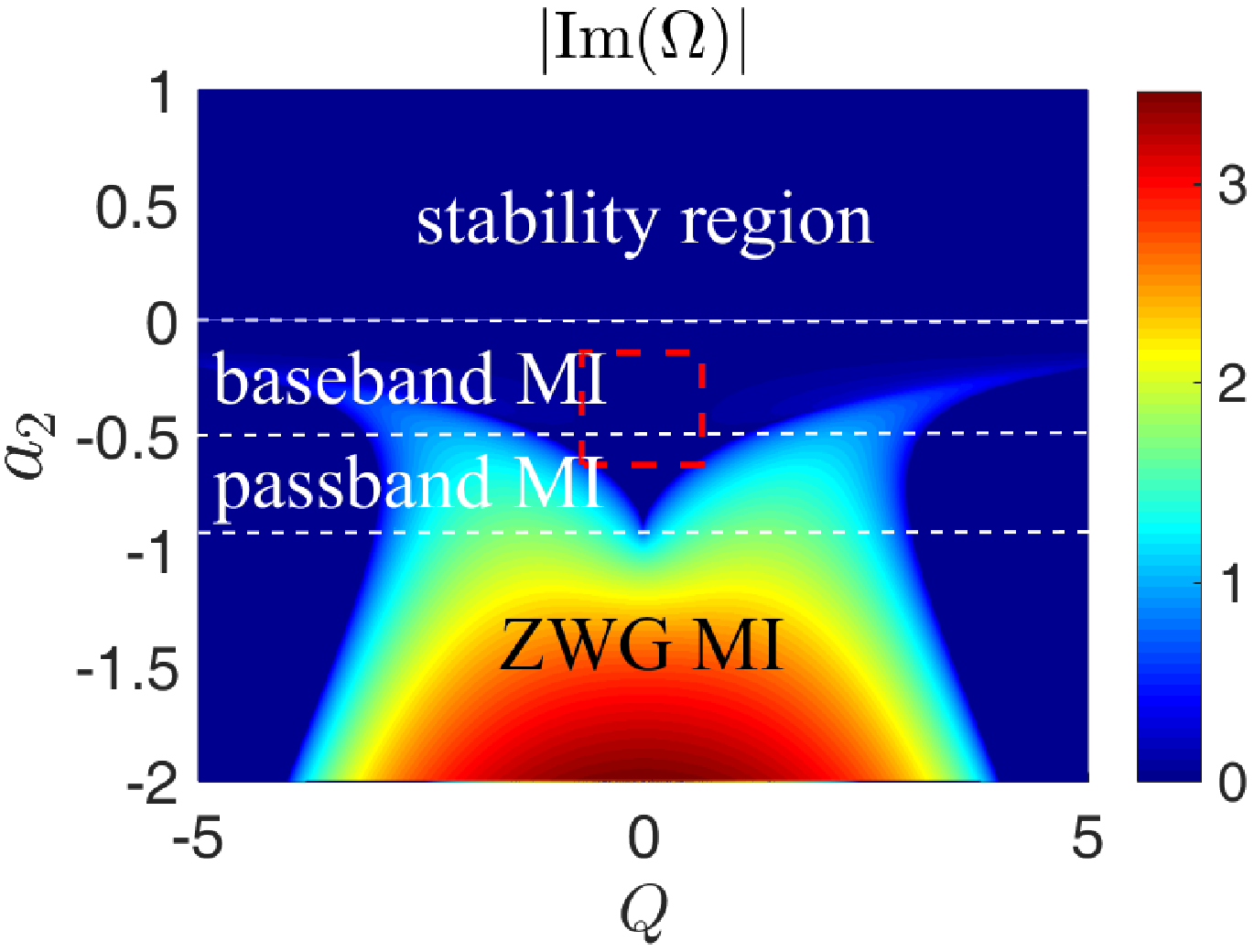}\hspace{0.6cm}%
\includegraphics[height=95pt,width=110pt]{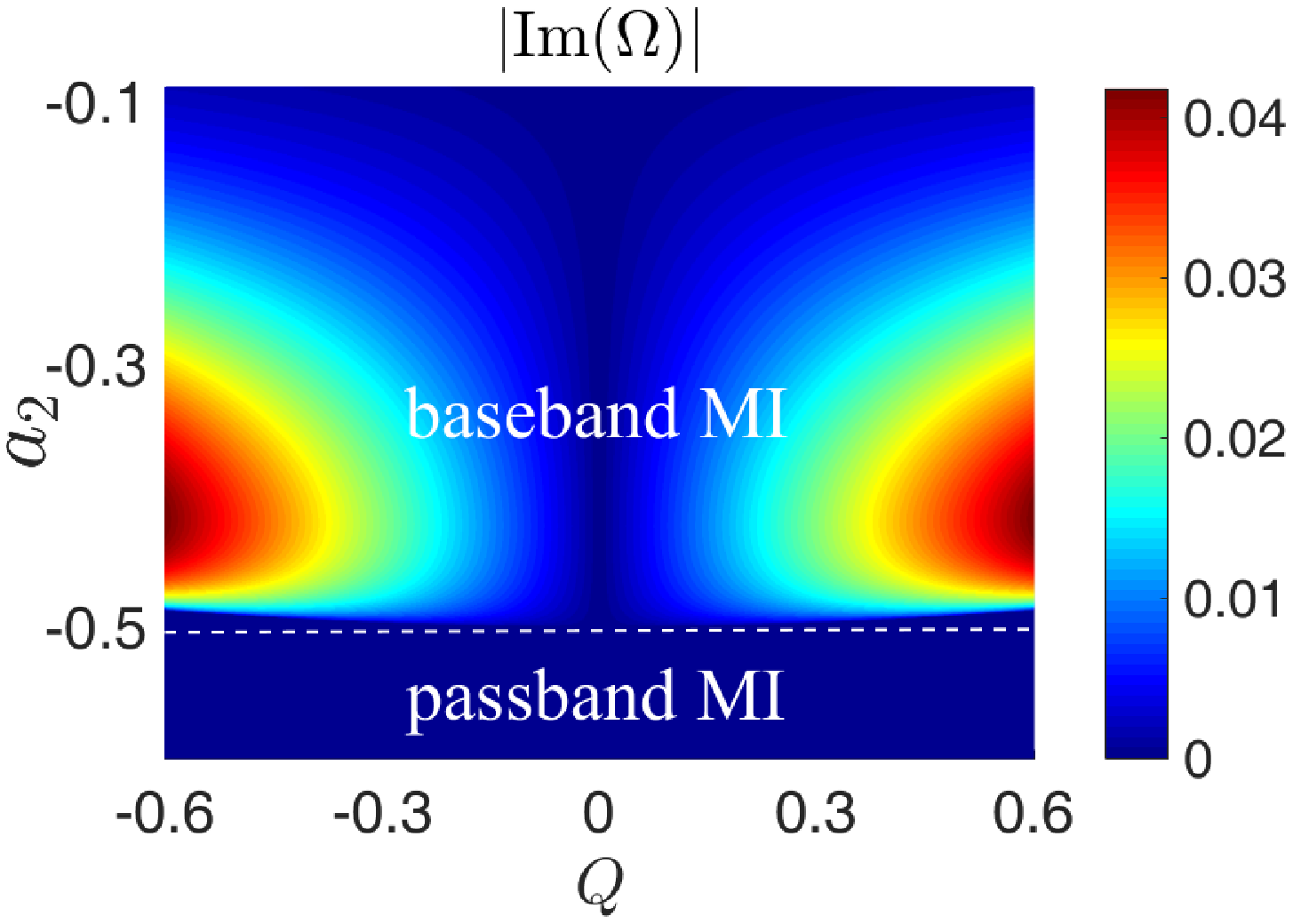} \newline
\vspace{-0.1cm}{\hspace{0cm}{\footnotesize (a)\hspace{4.1cm}(b)}}
\caption{The color map of the MI gain in parameter plane $(Q,a_{2})$ of CW
solutions (\protect\ref{dd1}), as produced by Eqs.~(\protect\ref{mt1}), with
fixed $a_{1}=2$. (b) Zoom of the red box in (a).}
\label{fig2}
\end{figure}

Figure~\ref{fig2} shows different MI types produced by Eqs.~(\ref{mt1}) with
fixed $a_{1}=2$. In particular, the modulational stability, baseband MI,
passband MI, and ZWG MI take place at $a_{2}>0$, $-0.5\leq a_{2}<0$, $%
-0.897<a_{2}<-0.5$, and $-31.7<a_{2}<-0.897$, respectively (at $a_{2}<-31.7$%
, the passband MI occurs, which is not shown in Fig.~\ref{fig2}). On the
other hand,~at $a_{1}=2$ Eq. (\ref{dd1}) produces the time-localized dark
modes solely in the last interval, $-31.7<a_{2}<-0.897$.

\section{Numerical simulations: Excitation of time-localized modes in
integrable and non-integrable MTM by chaotic perturbations added to the
background field}

The MI evolution is a natural source of solitary waves~\cite%
{Hasegawa,js1,js22}. In particular, the MI evolution initiated by random
perturbations has drawn interest in optics and hydrodynamics, chiefly in
connection to the generation of RWs and breathers~\cite{rw12,JNN2016}. To
verify the relation between the existence of the time-localized dark and
anti-dark modes and ZWG MI, we consider a possibility to excite such modes
from a chaotic background field in the presence of the ZWG-MI. For this
purpose, we simulated the evolution of the CW states taken as the initial
condition, perturbed by a random Gaussian noise of relative strength $5\%$.

\begin{figure}[tbp]
\centering
\includegraphics[height=85pt,width=113pt]{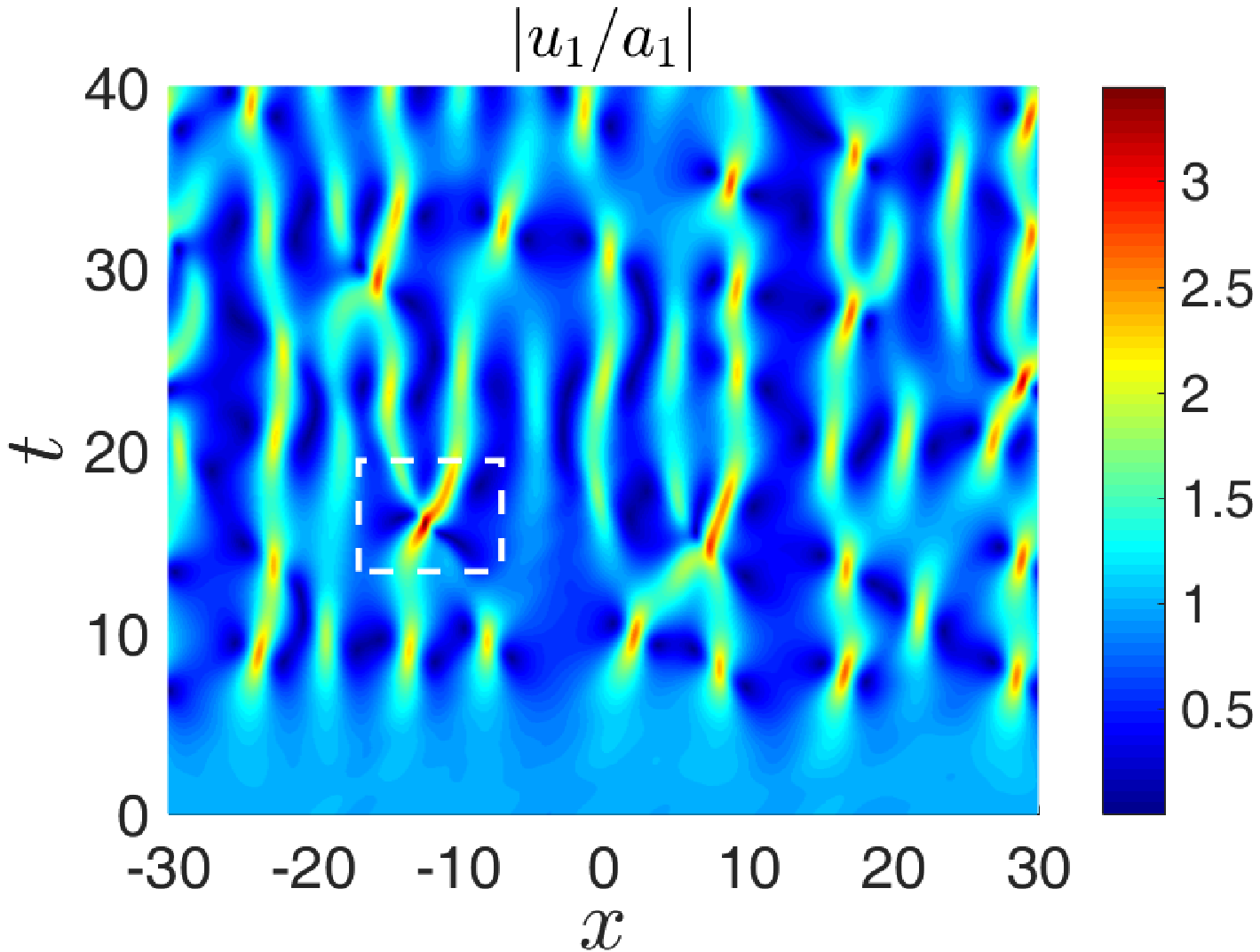}\hspace{0.6cm}%
\includegraphics[height=85pt,width=113pt]{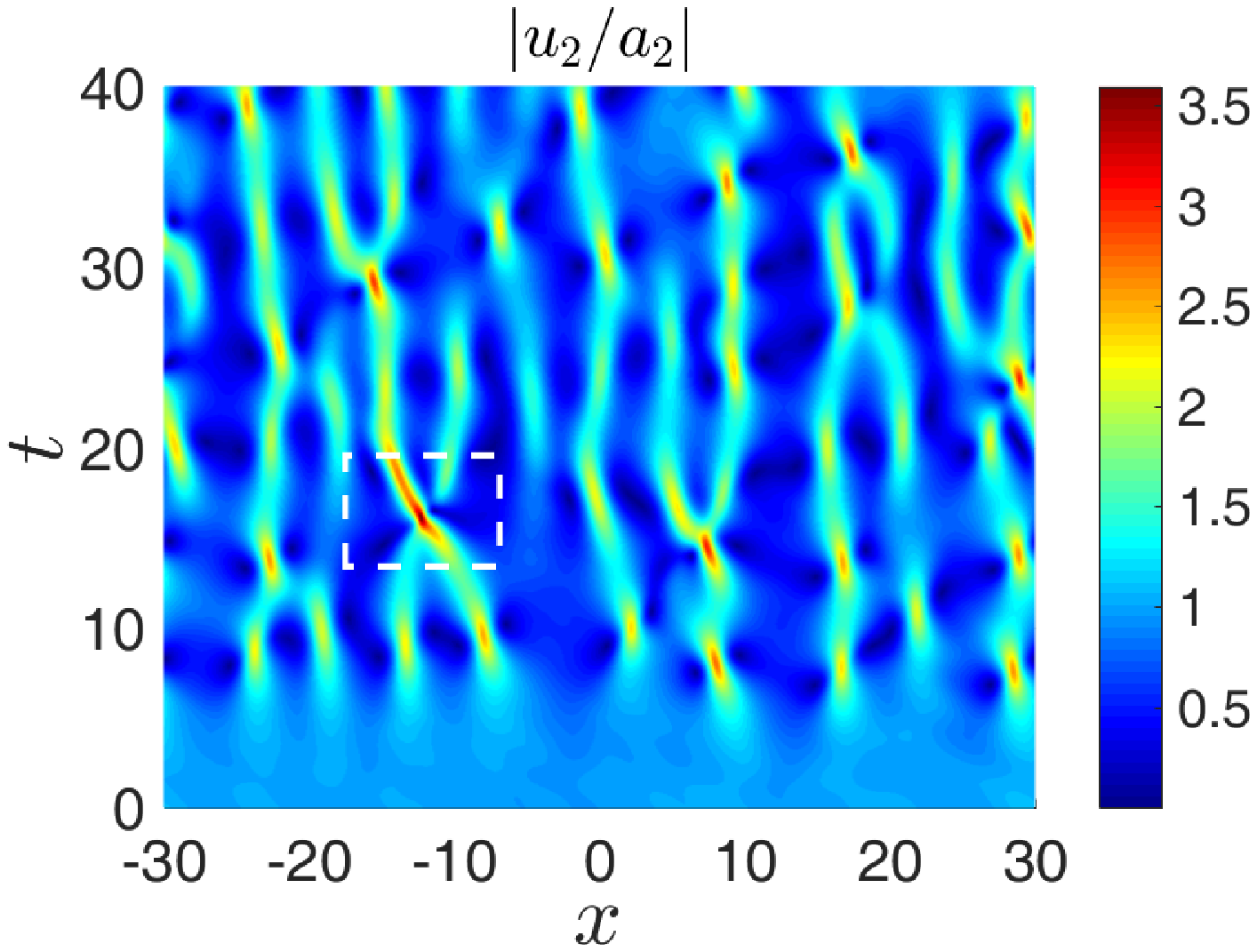} \newline
\vspace{0.2cm}{\hspace{0.9cm}{\footnotesize (a)\hspace{4cm}(b)}}\newline
\vspace{0.2cm} \includegraphics[height=85pt,width=113pt]{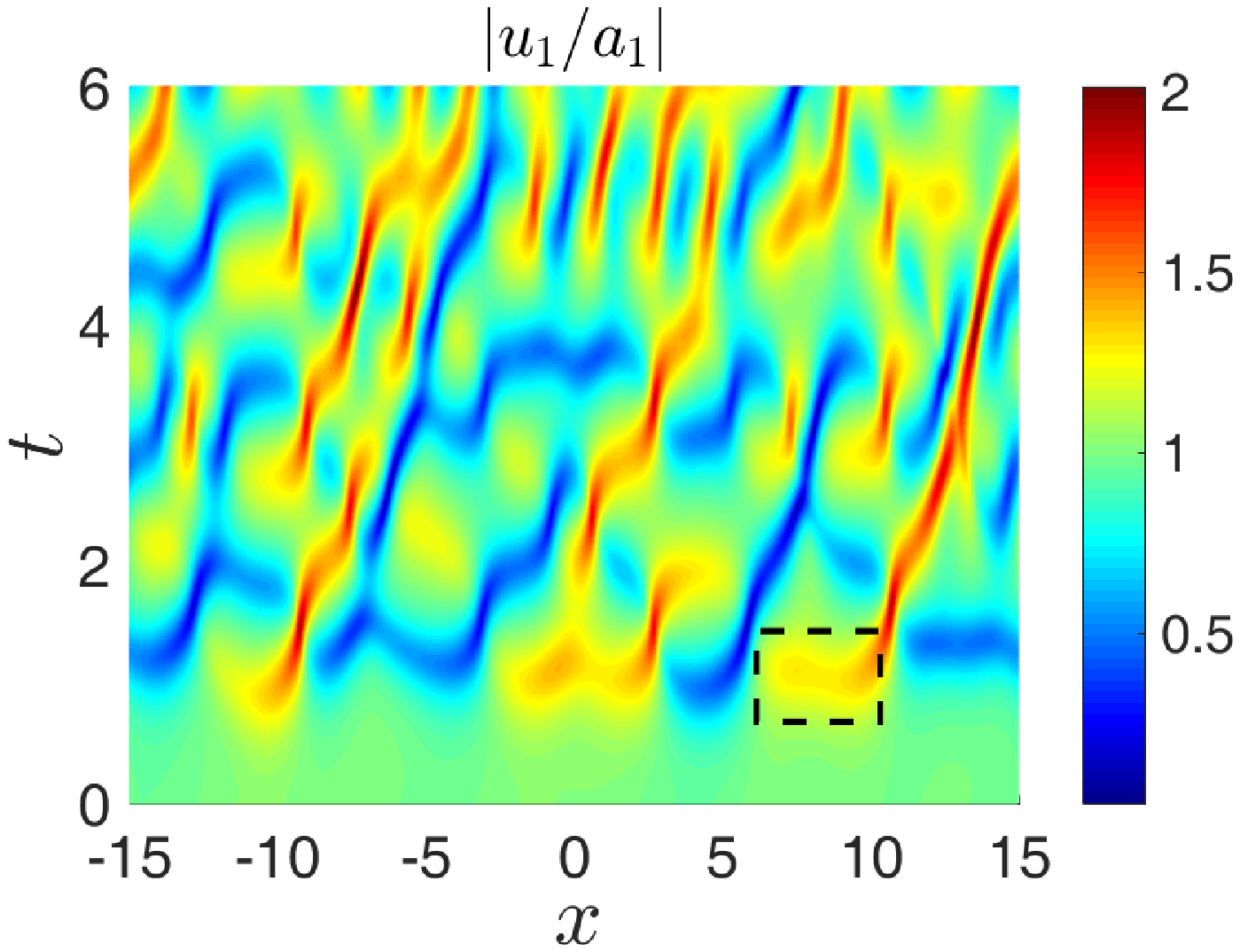}\hspace{%
0.6cm}\includegraphics[height=85pt,width=113pt]{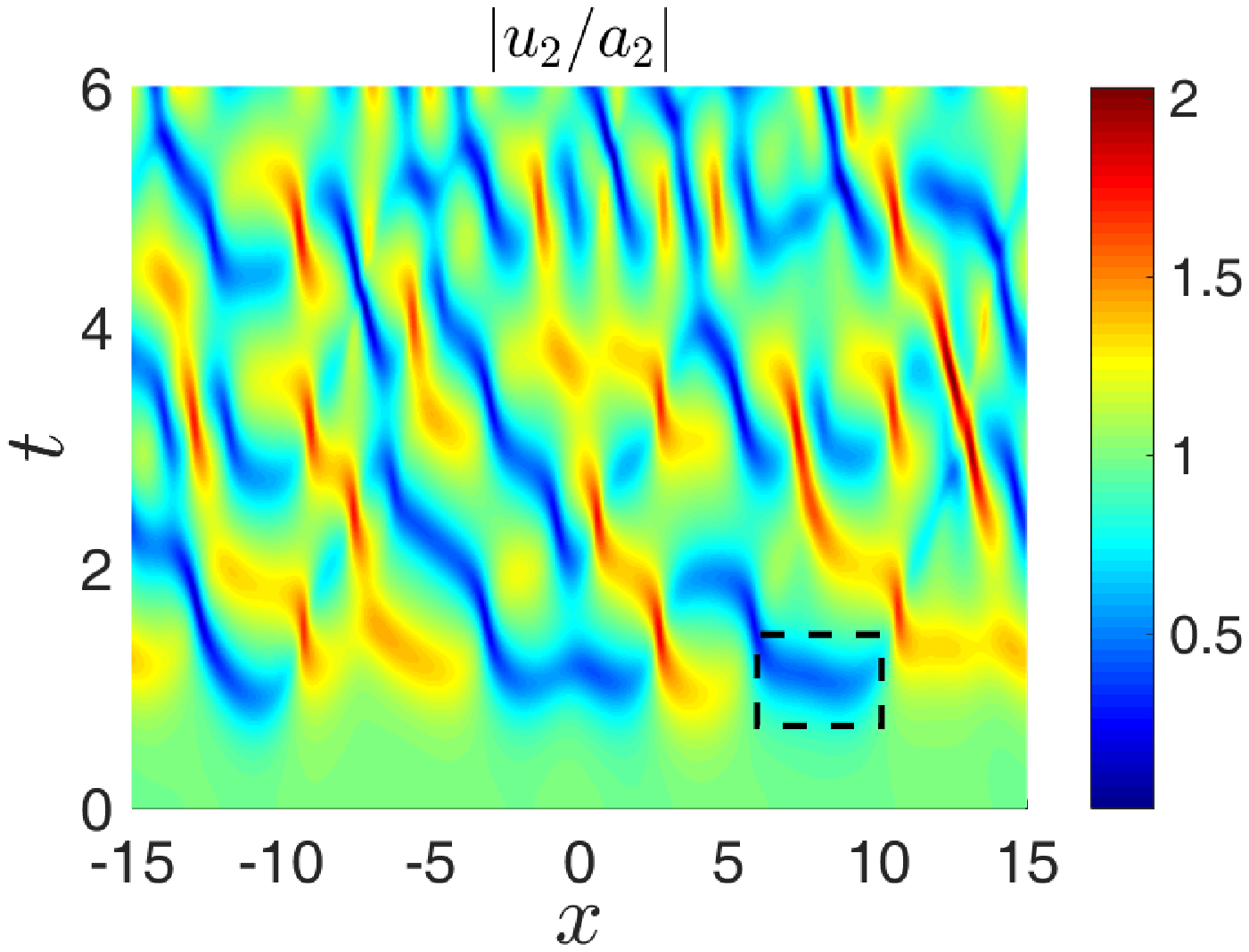} \newline
\vspace{0.2cm}{\hspace{0.9cm}{\footnotesize (c)\hspace{4cm}(d)}}\newline
\vspace{0.2cm} \includegraphics[height=85pt,width=113pt]{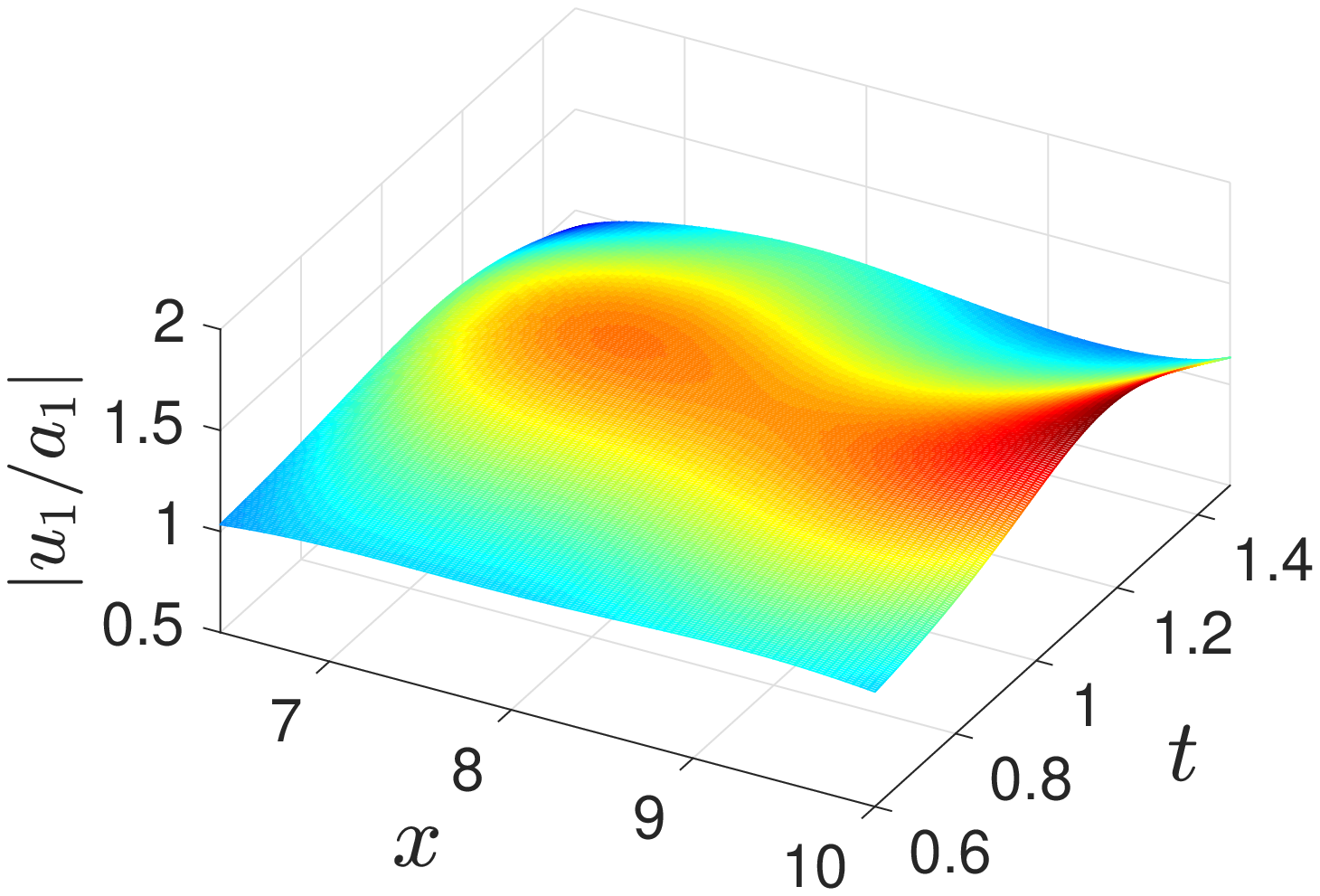}%
\hspace{0.6cm}\includegraphics[height=85pt,width=113pt]{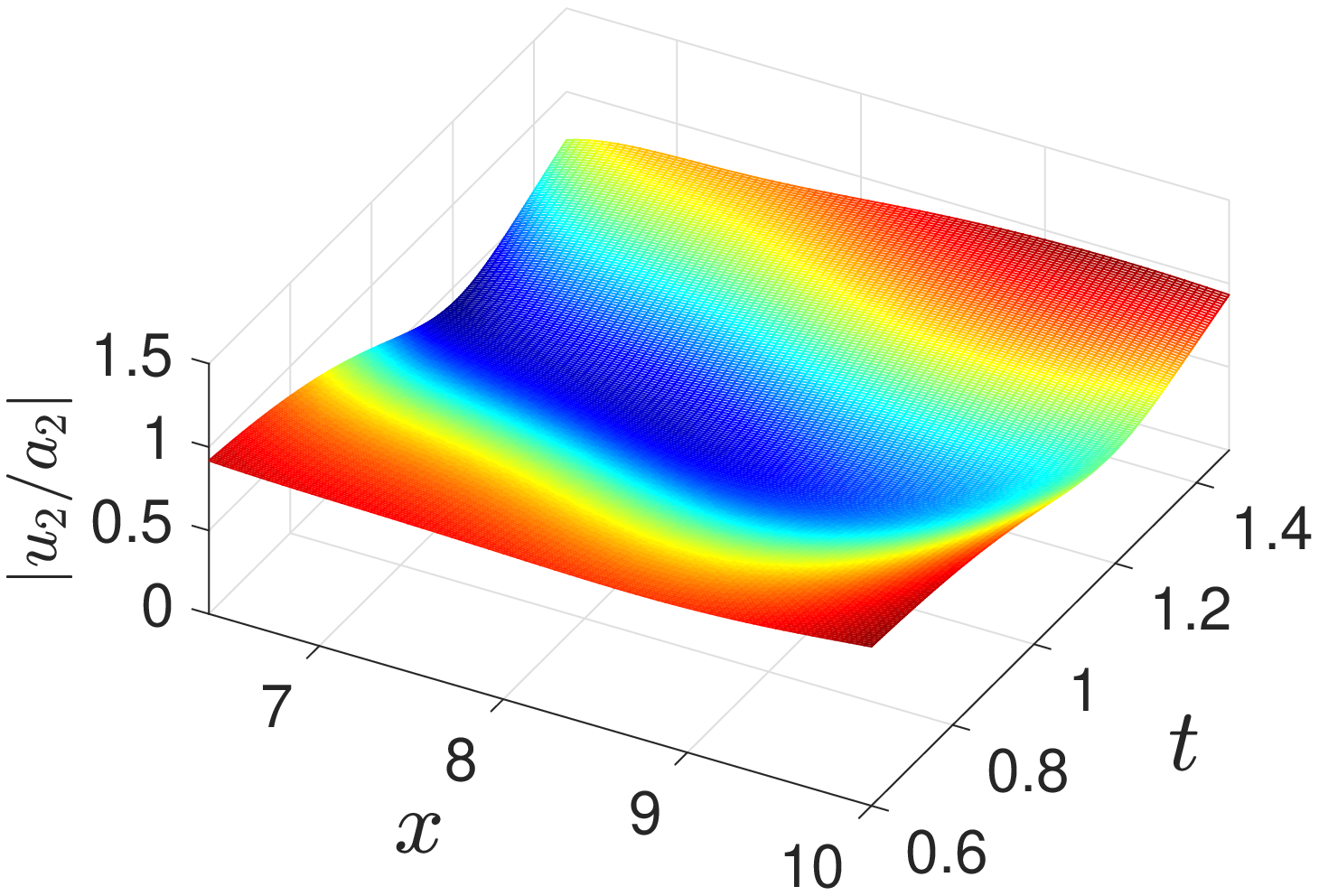}
\newline
\vspace{-0.2cm}{\hspace{-0cm}{\footnotesize (e)\hspace{4cm}(f)}}
\caption{The numerically simulated excitation of a pattern composed of
time-localized dark modes by chaotic perturbations with a $5\%$ relative
strength initially added to the CW\ background. The parameters are: $%
a_{1}=-a_{2}=0.8$ in (a,b), and $a_{1}=-a_{2}=2.4$ in (c,d). A particular
frament in the form of a dark-antidark localized mode is singled out by the
black box. Panels (e,f) display the three-dimensional zoom of this pattern.}
\label{fig3}
\end{figure}

As demonstrated in Fig.~\ref{fig3}, the noisy background features apparent
MI-driven chaotic dynamics. 
For parameters $a_{1}=-a_{2}=0.8$ in Figs.~\ref{fig3}(a,b), which satisfy
the RW existence condition~\cite{ASA2015,LWB2022,JBB2022}, but do not
satisfy condition (\ref{infconstraint4}) for the occurrence of the ZWG MI,
isolated peaks with amplitudes $\sim $ three times the background level
emerge at random positions. Actually, these are RWs, while no soliton-like
states appear in Figs.~\ref{fig3}(a,b).

On the other hand, for parameters $a_{1}=-a_{2}=2.4$, which satisfy
condition (\ref{infconstraint4}), the evolution initiated by the chaotic
perturbation produces localized soliton-like structures in Figs.~\ref{fig3}%
(c,d), while the peak amplitudes are less than twice the background level.
In particular, a structure which is recognized as a (portion of a)
time-localized mode with dark and anti-dark components, similar to that
displayed in Figs.~\ref{fig1}(c,d), is singled out by a black box in Figs.~%
\ref{fig3}(c,d). Further, Figs.~\ref{fig3}(e,f) show enlarged
three-dimensional plots of this wave pattern.

\begin{figure}[tbp]
\centering
\includegraphics[height=85pt,width=113pt]{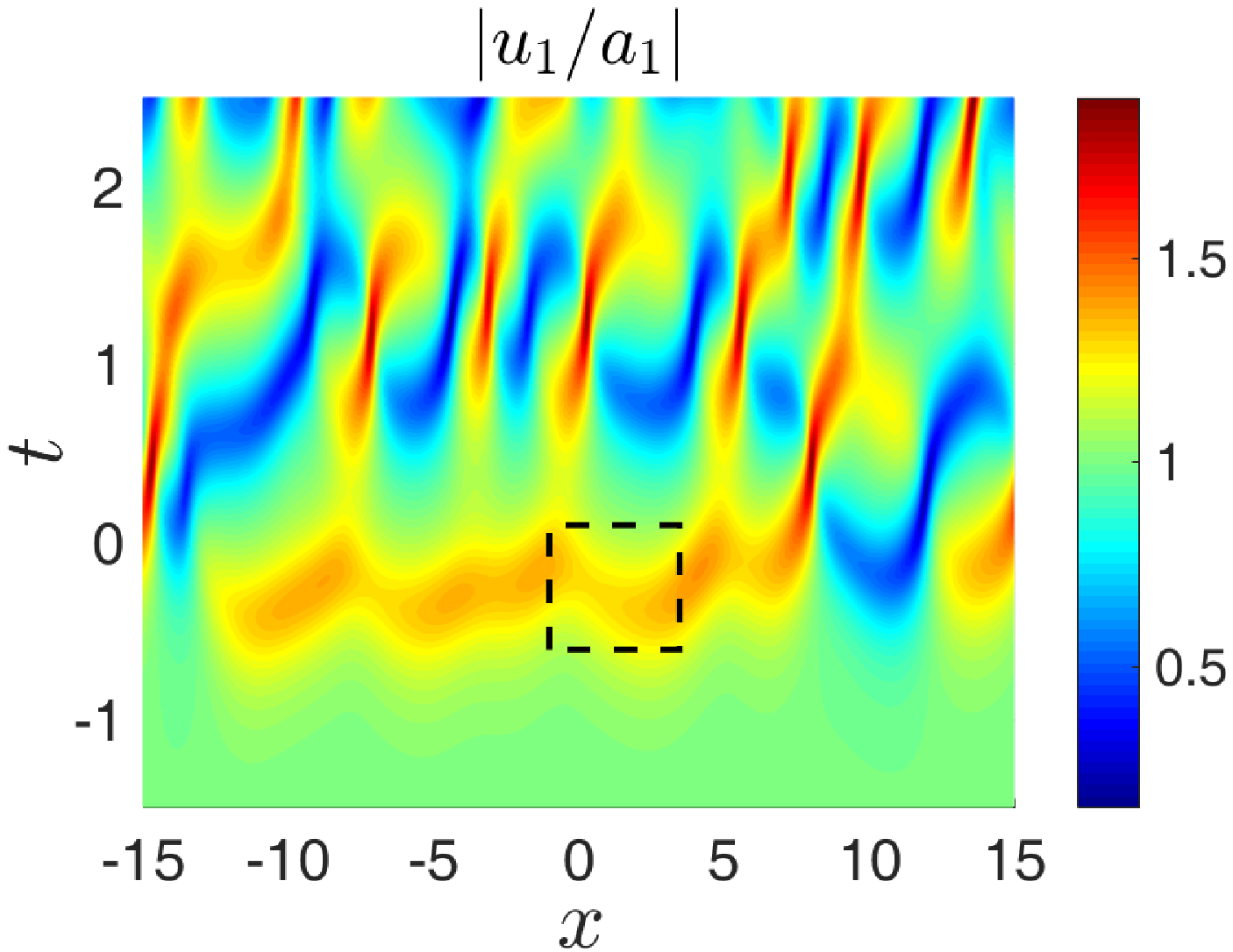}\hspace{0.6cm}%
\includegraphics[height=85pt,width=113pt]{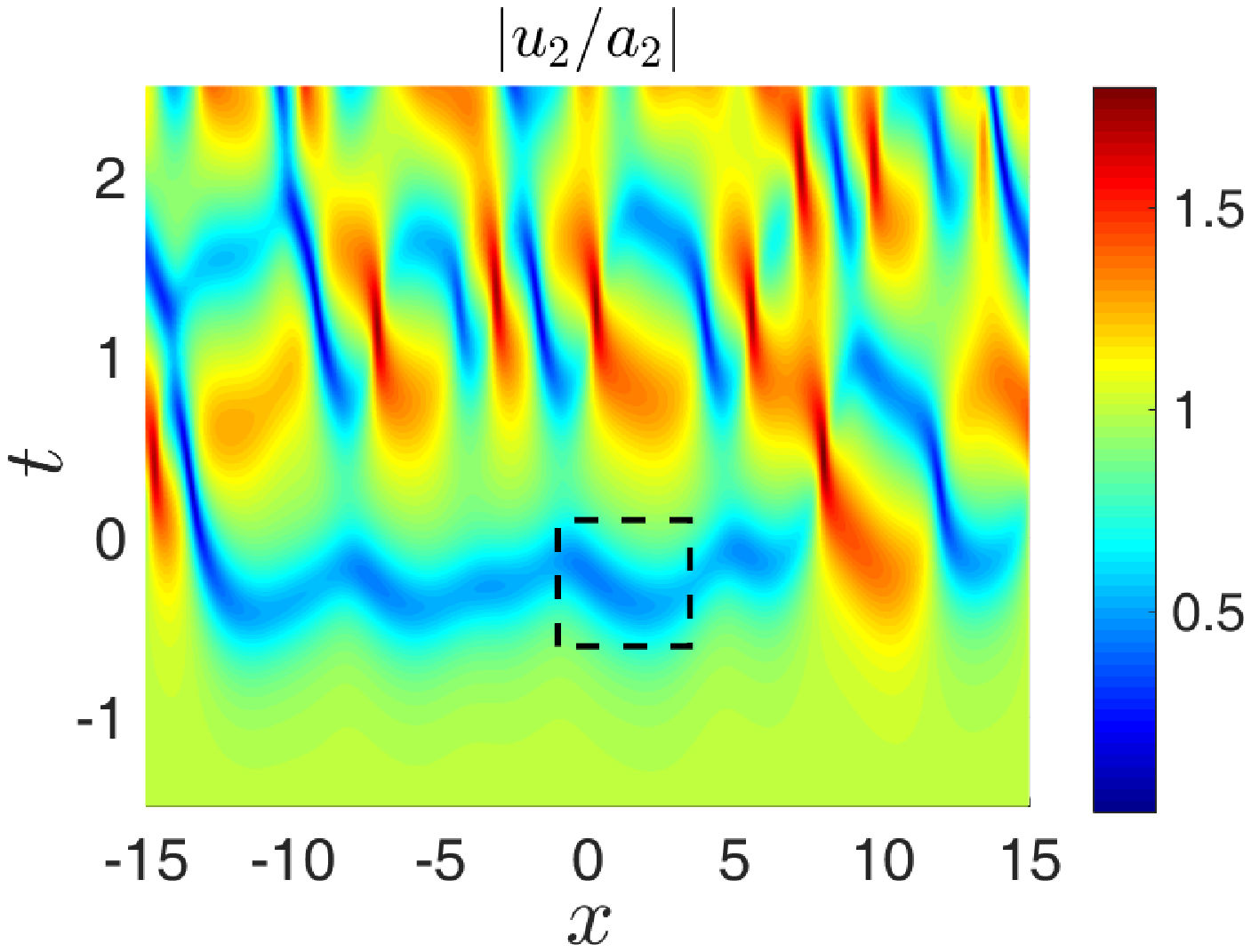} \newline
\vspace{0.2cm}{\hspace{0.9cm}{\footnotesize (a)\hspace{4cm}(b)}}\newline
\vspace{0.2cm} \includegraphics[height=85pt,width=113pt]{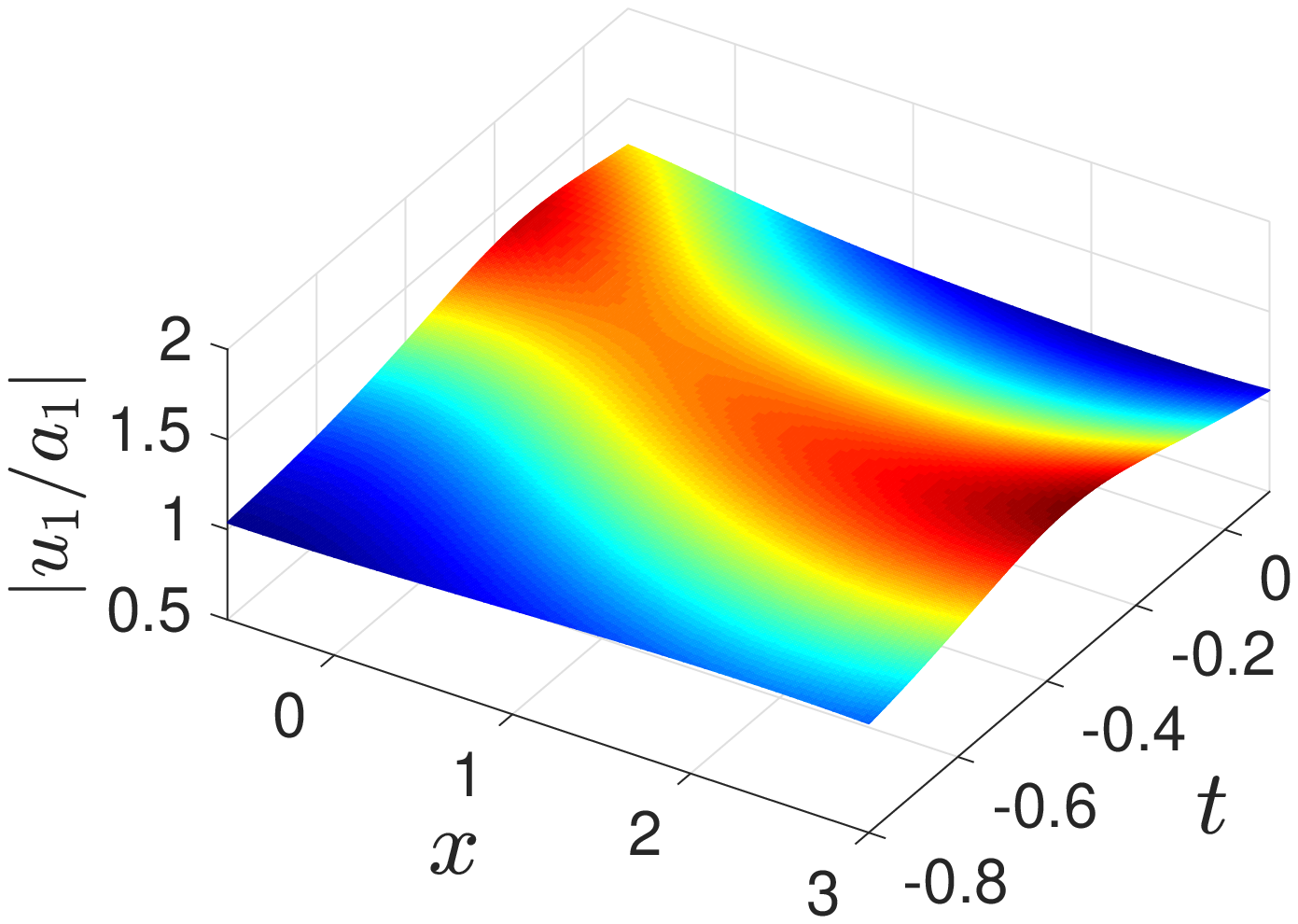}%
\hspace{0.6cm}\includegraphics[height=85pt,width=113pt]{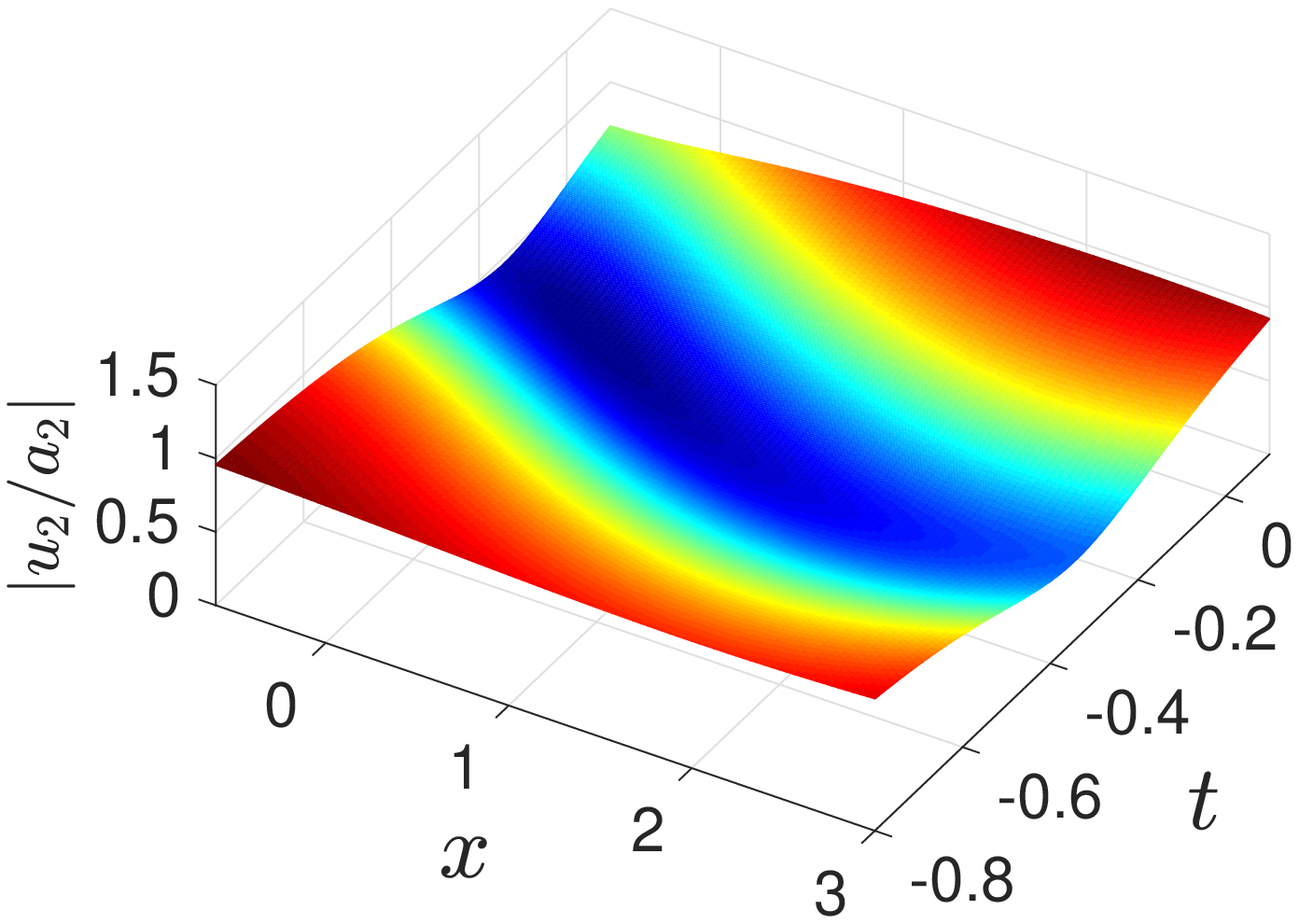}
\newline
\vspace{-0.2cm}{\hspace{-0cm}{\footnotesize (c)\hspace{4cm}(d)}}
\caption{The simulated evolution of a time-localized mode given by
solution~(\protect\ref{dd1}) with initially added random-noise perturbations
at the $2\%$ level. The parameters are $\protect\beta =1$, $\protect\xi %
^{(0)}=0$, $a_{1}=-a_{2}=2.4$ and $p_{1}=\protect\sqrt{119}/3$. The
numerical simulation is initiated at $t=-1.5$. A particular fragment of a
dark mode is singled out by a black box. Panels (c,d) display the
three-dimensional zoom of this pattern.}
\label{fig4}
\end{figure}

Similar to RWs, the time-localized dark modes are 
sensitive to the presence of perturbations, because their background is
subject to MI. Figure~\ref{fig4} exhibits the evolution of the
time-localized mode with initially added $2\%$ random Gaussian-noise
perturbations. It is observed that, although the quasi-soliton pattern is
affected by the background instability, fragments of the time-localized dark
state, which are also localized in the $x$ direction, persist as robust
elements of the emerging complex pattern, as shown in Figs.~\ref{fig4}(c,d)
by the three-dimensional zoom of the fragment singled out by the black box
in Figs. \ref{fig4}(a,b). Note that Figs.~\ref{fig3}(e,f) and Figs.~\ref%
{fig4}(c,d) exhibit similar coupled dark-antidark structures, implying that,
in Figs.~\ref{fig3}(c,d), the ZWG MI indeed produces complex patterns
incorporating time-localized modes. 

\begin{figure}[tbp]
\centering
\includegraphics[height=85pt,width=113pt]{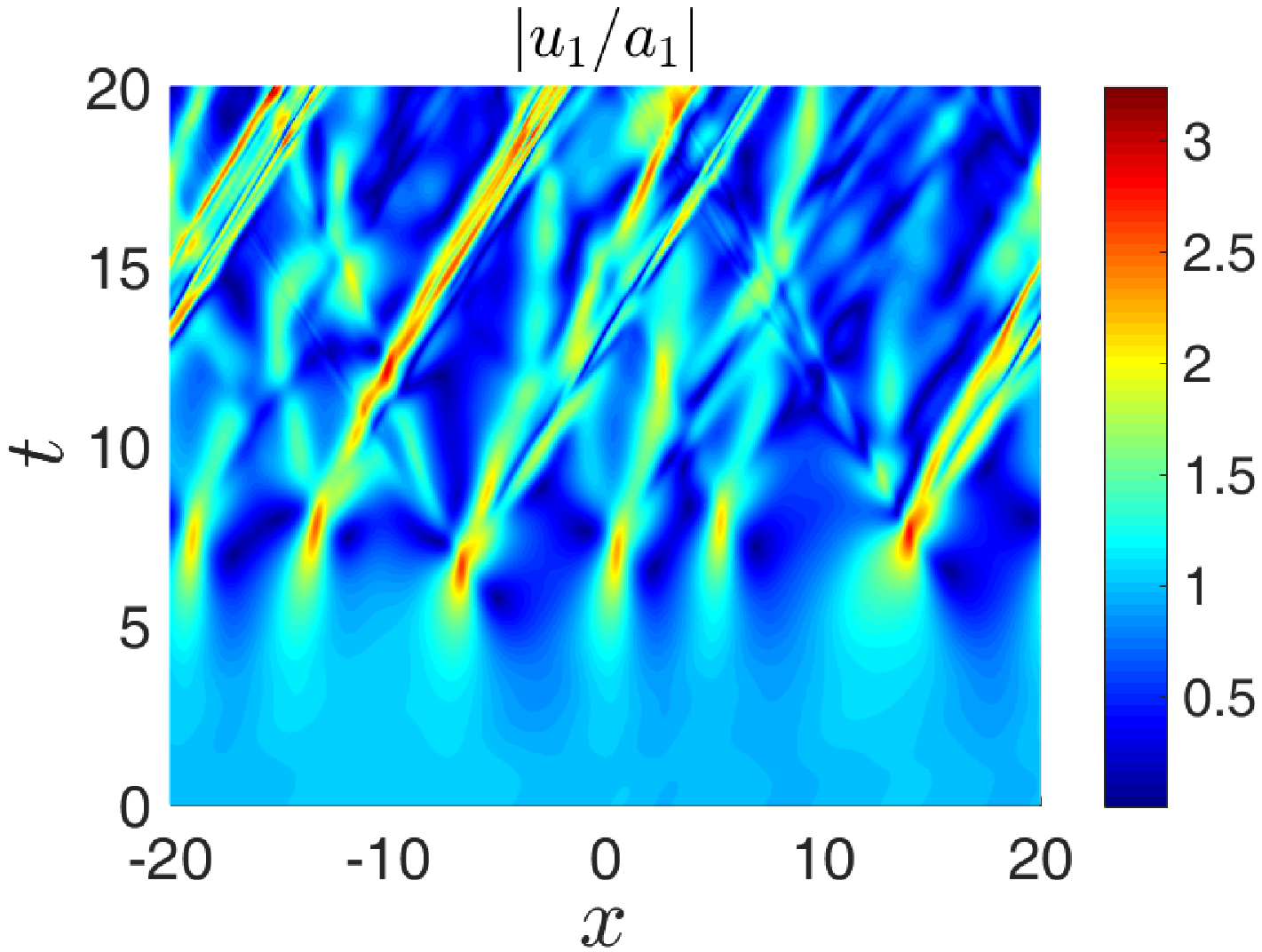}\hspace{0.6cm}%
\includegraphics[height=85pt,width=113pt]{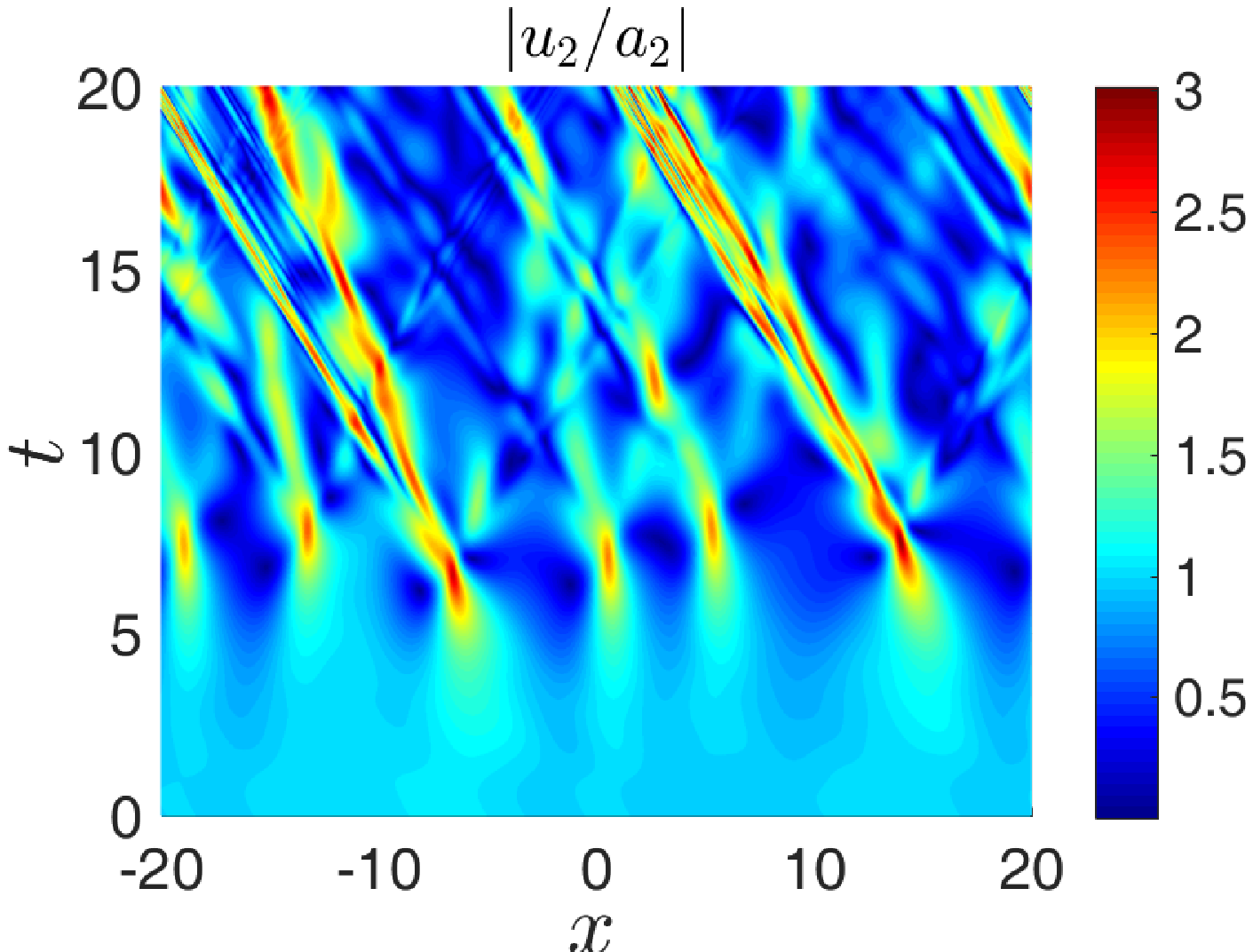} \newline
\vspace{0.2cm}{\hspace{0.9cm}{\footnotesize (a)\hspace{4cm}(b)}}\newline
\vspace{0.2cm} \includegraphics[height=85pt,width=113pt]{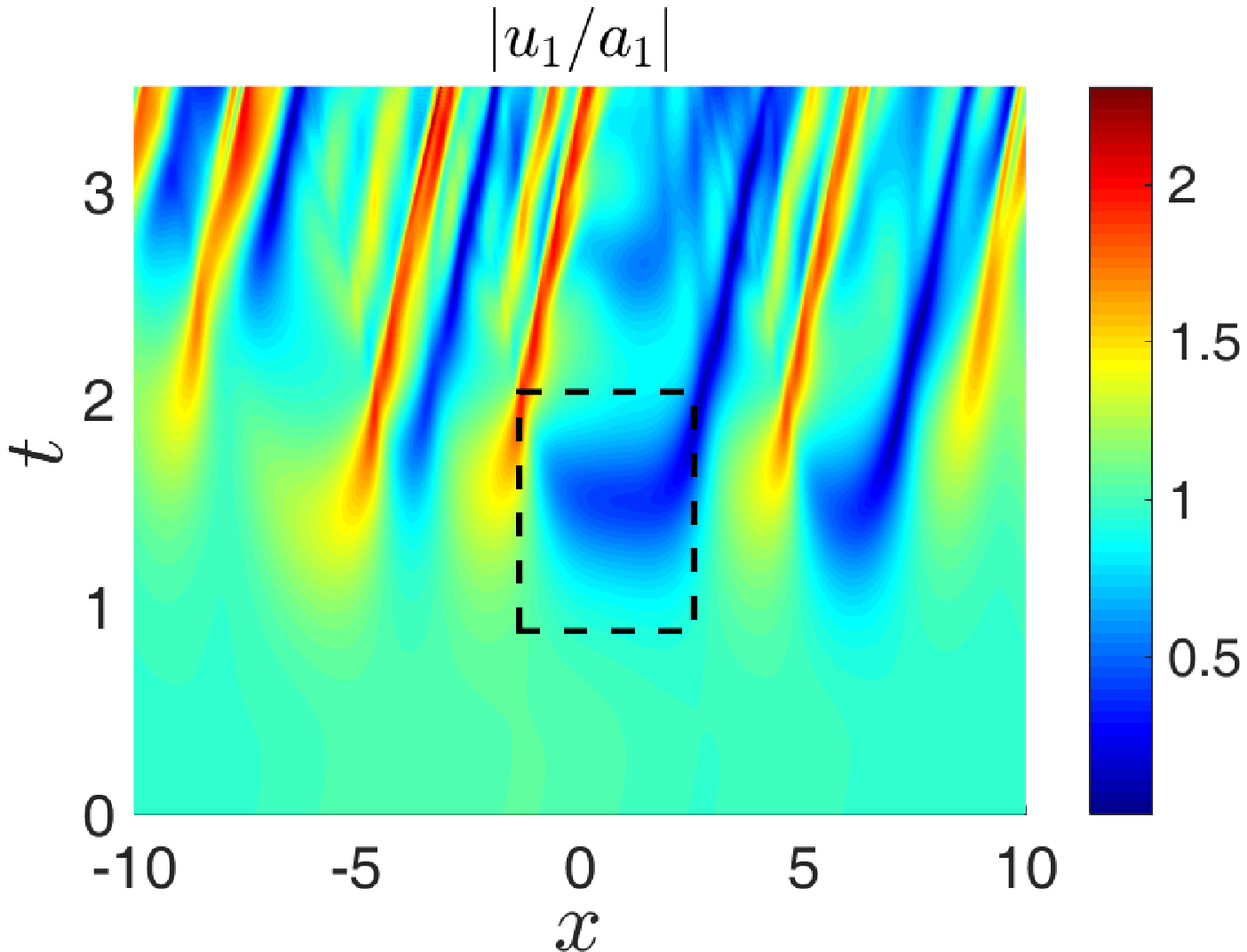}\hspace{%
0.6cm}\includegraphics[height=85pt,width=113pt]{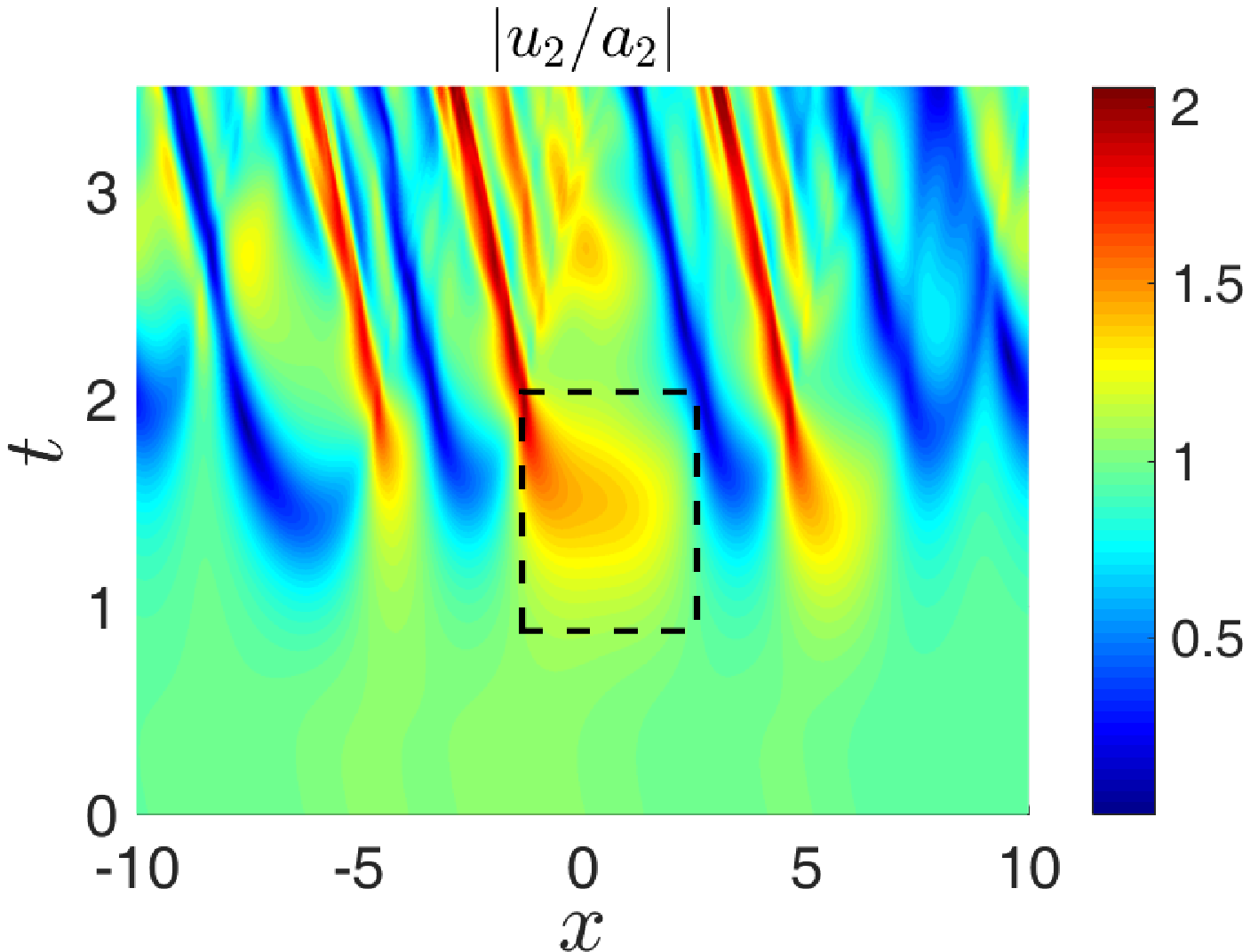} \newline
\vspace{0.2cm}{\hspace{0.9cm}{\footnotesize (c)\hspace{4cm}(d)}}\newline
\vspace{0.2cm} \includegraphics[height=85pt,width=113pt]{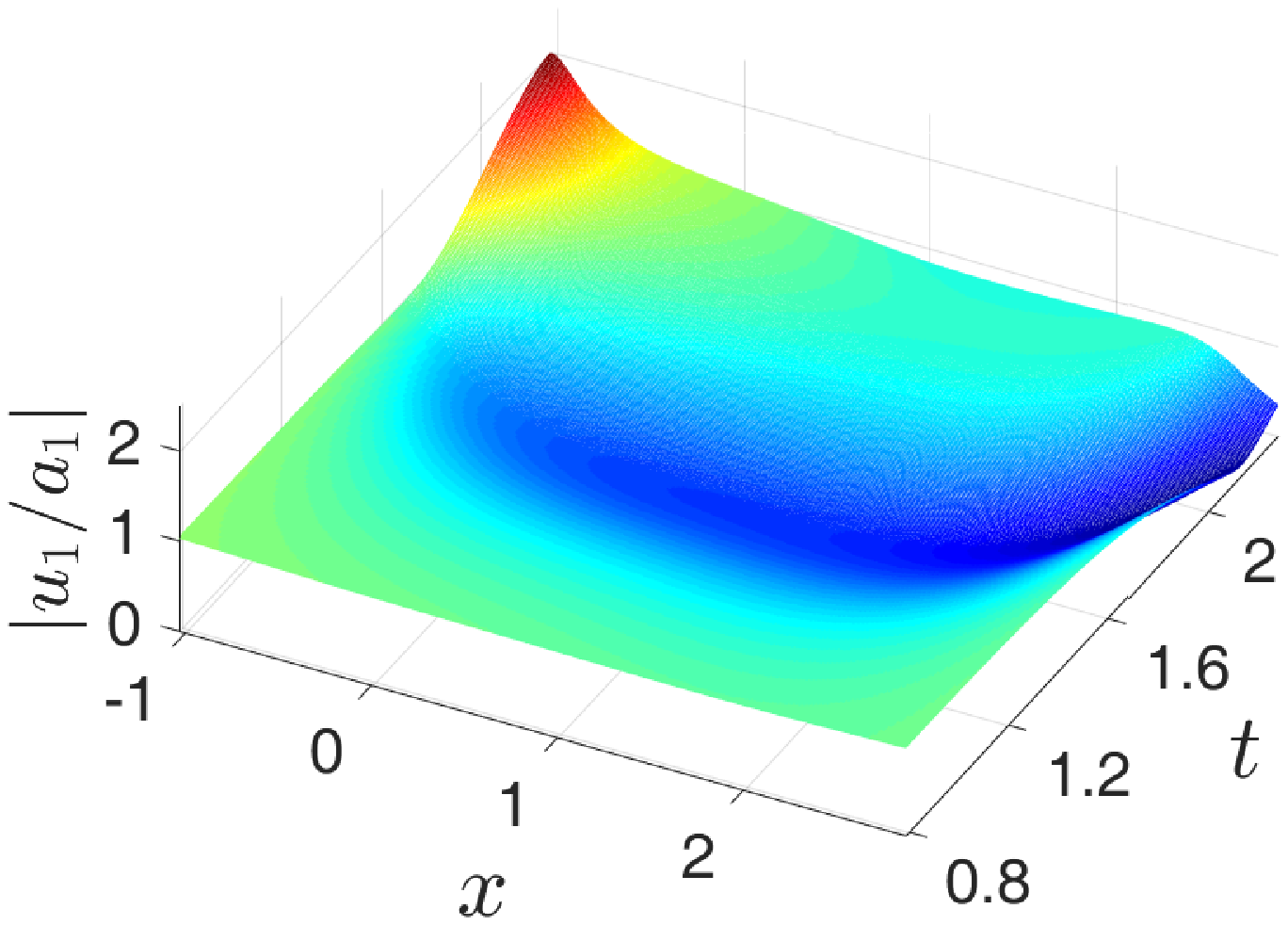}\hspace{%
0.6cm}\includegraphics[height=85pt,width=113pt]{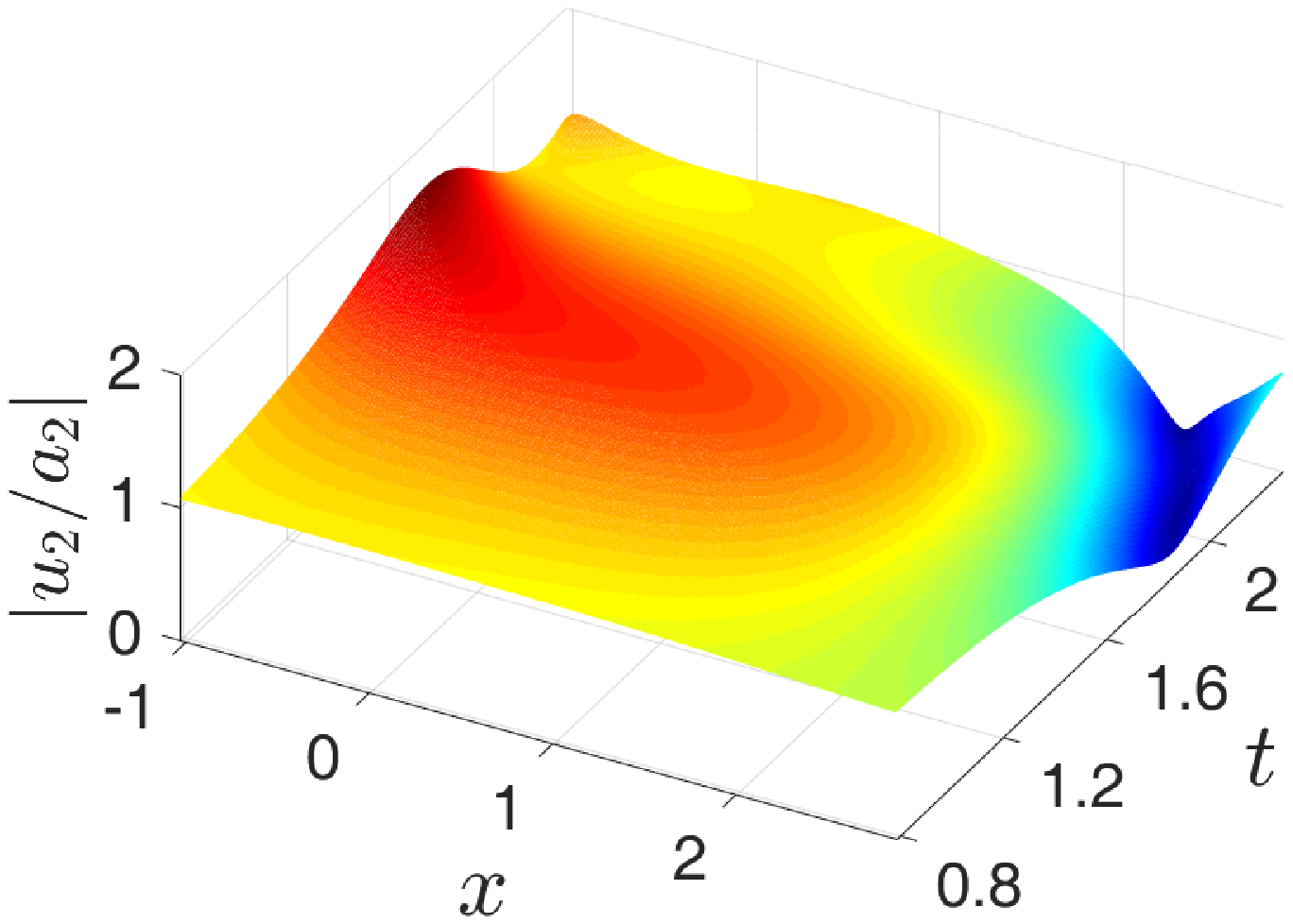} \newline
\vspace{-0.2cm}{\hspace{-0cm}{\footnotesize (e)\hspace{4cm}(f)}}
\caption{The excitation of a pattern composed of time-localized dark modes,
as produced by simulations of Eq. (\protect\ref{mt123}) with $\protect\gamma %
=0.5$. The input is the CW background perturbed by a random noise with the $%
5\%$ strength. The parameters are $a_{1}=-a_{2}=0.8$ (corresponding to the
baseband MI) in (a,b), and $a_{1}=-a_{2}=2.4$ (the ZWG MI) in (c,d). A
particular time-localized dark mode is singled out by the black box. Panels
(e,f) display the three-dimensional zoom of this state.}
\label{fig5}
\end{figure}

It is quite interesting to find time-localized dark (and antidark) modes as
solutions of the coupled-mode equations (the non-integrable version of MTM)
which furnish, as mentioned above, a model for the light propagation in
periodic or Bragg nonlinear optical media. The respective non-integrable
extension of Eq. (\ref{mt1}) is
\begin{subequations}
\label{mt123}
\begin{eqnarray}
&&i\partial _{t}u_{1}+i\partial _{x}u_{1}+u_{2}+\left( |u_{2}|^{2}+\gamma
|u_{1}|^{2}\right) u_{1}=0, \\
&&i\partial _{t}u_{2}-i\partial _{x}u_{2}+u_{1}+\left( |u_{1}|^{2}+\gamma
|u_{2}|^{2}\right) u_{2}=0.
\end{eqnarray}%
It differs from the integrable MTM by the presence of the SPM (self-phase
modulation) with relative strength $\gamma $. A straightforward extension of
the above analysis produces the following existence condition for the ZWG-MI
in the present case:
\end{subequations}
\begin{equation}
2+\frac{a_{1}^{2}}{a_{2}^{2}}+\frac{a_{2}^{2}}{a_{1}^{2}}+4(1-\gamma
)a_{1}a_{2}<0,
\end{equation}%
cf. Eq. (\ref{infconstraint4}).\ For example, for the physically relevant
case of $\gamma =0.5$, Fig.~\ref{fig5} displays patterns which are quite
similar to those in Figs.~\ref{fig3}. This result confirms that the ZWG-MI
mechanism of the creation of the time-localized modes naturally extends to
the physically relevant non-integrable system and produces an experimentally
available setting where such states may be created. It is also relevant to
mention bright time-localized modes, which were very recently predicted
as solutions of Eq. (\ref{mt123})~\cite{Segev}. Unlike the
present considerations, those bright temporal
waves are not related to the CW background and MI conditions.

\section{The three-wave resonant-interaction system}

To demonstrate that the mechanism elaborated above can be readily
implemented in other systems, we consider the system for complex amplitudes $%
E_{n}=E_{n}(x,t)$ ($n=1,2,3$) of three waves coupled by the quadratic
interactions:
\begin{subequations} \label{threeequ}
\begin{eqnarray}
&&\partial _{t}E_{1}+V_{1}\partial _{x}E_{1}=\sigma _{1}E_{2}^{\ast
}E_{3}^{\ast }, \\
&&\partial _{t}E_{2}+V_{2}\partial _{x}E_{2}=\sigma _{2}E_{1}^{\ast
}E_{3}^{\ast }, \\
&&\partial _{t}E_{3}+V_{3}\partial _{x}E_{3}=\sigma _{3}E_{1}^{\ast
}E_{2}^{\ast }. 
\end{eqnarray}%
Here, $V_{n}$ are group velocities of the components, and $\sigma _{n}=\pm 1$
are signs of the interactions, which correspond to the
stimulated-backscattering ($\sigma _{1}=\sigma _{2}=-\sigma _{3}=1$ or $%
\sigma _{1}=-\sigma _{2}=-\sigma _{3}=1$), explosive ($\sigma _{1}=\sigma
_{2}=\sigma _{3}=1$), or soliton-exchange ($\sigma _{1}=-\sigma _{2}=\sigma
_{3}=1$) regime. As a fundamental model, system~(\ref{threeequ}) describes
diverse physical contexts in hydrodynamics, optics and plasmas~\cite%
{Kaup,3w2,3w1}. Without loss of generality, we set $V_{1}>V_{2}>V_{3}\equiv
0 $, in the reference frame co-moving with wave $E_{3}$.

It is well-known that system~(\ref{threeequ}) is completely integrable~\cite%
{Kaup, MFA2011,GZX2018}. The bilinear form~\cite{BJ2021} of system~(\ref%
{threeequ}) (the Hirota method) produces the fundamental three-component
dark-mode solutions admitted by the integrable system:
\end{subequations}
\begin{subequations}
\begin{gather}
E_{1}=\rho _{1}e^{i\phi _{1}}\frac{1-\frac{1}{p_{1}+p_{1}^{\ast }}\frac{%
p_{1}-i}{p_{1}^{\ast }+i}e^{\eta _{1}+\eta _{1}^{\ast }}}{1+\frac{1}{%
p_{1}+p_{1}^{\ast }}e^{\eta _{1}+\eta _{1}^{\ast }}}, \\
E_{2}=\rho _{2}e^{i\phi _{2}}\frac{1-\frac{1}{p_{1}+p_{1}^{\ast }}\frac{%
p_{1}^{\ast }}{p_{1}}e^{\eta _{1}+\eta _{1}^{\ast }}}{1+\frac{1}{%
p_{1}+p_{1}^{\ast }}e^{\eta _{1}+\eta _{1}^{\ast }}}, \\
\hspace{-1.5cm}E_{3}=i\rho _{3}e^{-i(\phi _{1}+\phi _{2})}\frac{1+\frac{1}{%
p_{1}+p_{1}^{\ast }}\frac{p_{1}^{\ast }+i}{p_{1}-i}\frac{p_{1}}{p_{1}^{\ast }%
}e^{\eta _{1}+\eta _{1}^{\ast }}}{1+\frac{1}{p_{1}+p_{1}^{\ast }}e^{\eta
_{1}+\eta _{1}^{\ast }}},
\end{gather}%
where
\end{subequations}
\begin{gather*}
\phi _{l}=c_{l}x+d_{l}t,\ (l=1,2),d_{1}=d_{2}=\frac{\gamma _{3}}{2}, \\
c_{1,2}=-\frac{2\gamma _{1,2}+\gamma _{3}}{2V_{1,2}}, \\
\eta _{1}=\frac{1}{p_{1}}r+\frac{1}{p_{1}-i}s+\eta _{1}^{(0)}, \\
r=\frac{\gamma _{1}}{V_{1}-V_{2}}(x-V_{2}t), \\
s=\frac{\gamma _{2}}{V_{2}-V_{1}}(x-V_{1}t).
\end{gather*}%
Here $\rho _{n}$ are nonzero real constants representing the background
amplitudes of the dark-soliton components $E_{n}$, $p_{1}$ and $\eta
_{1}^{(0)}$ are complex constants,
\begin{equation}
\gamma _{1}=\sigma _{1}\frac{\rho _{2}\rho _{3}}{\rho _{1}},\gamma
_{2}=\sigma _{2}\frac{\rho _{1}\rho _{3}}{\rho _{2}},\gamma _{3}=\sigma _{3}%
\frac{\rho _{1}\rho _{2}}{\rho _{3}},
\end{equation}%
and these parameters satisfy the following constraint:
\begin{equation}
\frac{\gamma _{1}V_{2}}{|p_{1}|^{2}\gamma _{3}(V_{2}-V_{1})}-\frac{\gamma
_{2}V_{1}}{|p_{1}-i|^{2}\gamma _{3}(V_{2}-V_{1})}=1.  \label{threedd2}
\end{equation}

To cast the exact solution of system~(\ref{threeequ}) in the form of a
time-localized mode, we set
\begin{equation}
\text{Re}\left\{ \frac{1}{p_{1}}\frac{\gamma _{1}}{V_{1}-V_{2}}+\frac{1}{%
p_{1}-i}\frac{\gamma _{2}}{V_{2}-V_{1}}\right\} =0,
\end{equation}%
which yields
\begin{equation}
|p_{1}-i|^{2}\gamma _{1}-|p_{1}|^{2}\gamma _{2}=0.  \label{threeconstraint1}
\end{equation}%
Combining~Eqs. (\ref{threedd2}) and~(\ref{threeconstraint1}) and setting $%
p_{1}=p_{1R}+ip_{1I}$, we obtain
\begin{equation}
p_{1R}=\pm \frac{\sqrt{4\gamma _{1}\gamma _{2}-(\gamma _{1}+\gamma
_{2}-\gamma _{3})^{2}}}{2\gamma _{3}},\ p_{1I}=\frac{\gamma _{1}-\gamma
_{2}+\gamma _{3}}{2\gamma _{3}}.
\end{equation}
As $p_{1R}$ takes nonzero real values, parameters $\gamma _{1}$, $\gamma
_{2} $ and $\gamma _{3}$ need to satisfy the constraint
\begin{equation}
(\gamma _{1}+\gamma _{2}-\gamma _{3})^{2}-4\gamma _{1}\gamma _{2}<0.
\label{threeconstraint2}
\end{equation}%
Thus, Eq.~(\ref{threeconstraint2}) is the existence condition for the
time-localized dark modes as solutions of system~(\ref{threeequ}).

\begin{figure}[tbp]
\centering
\includegraphics[height=100pt,width=80pt]{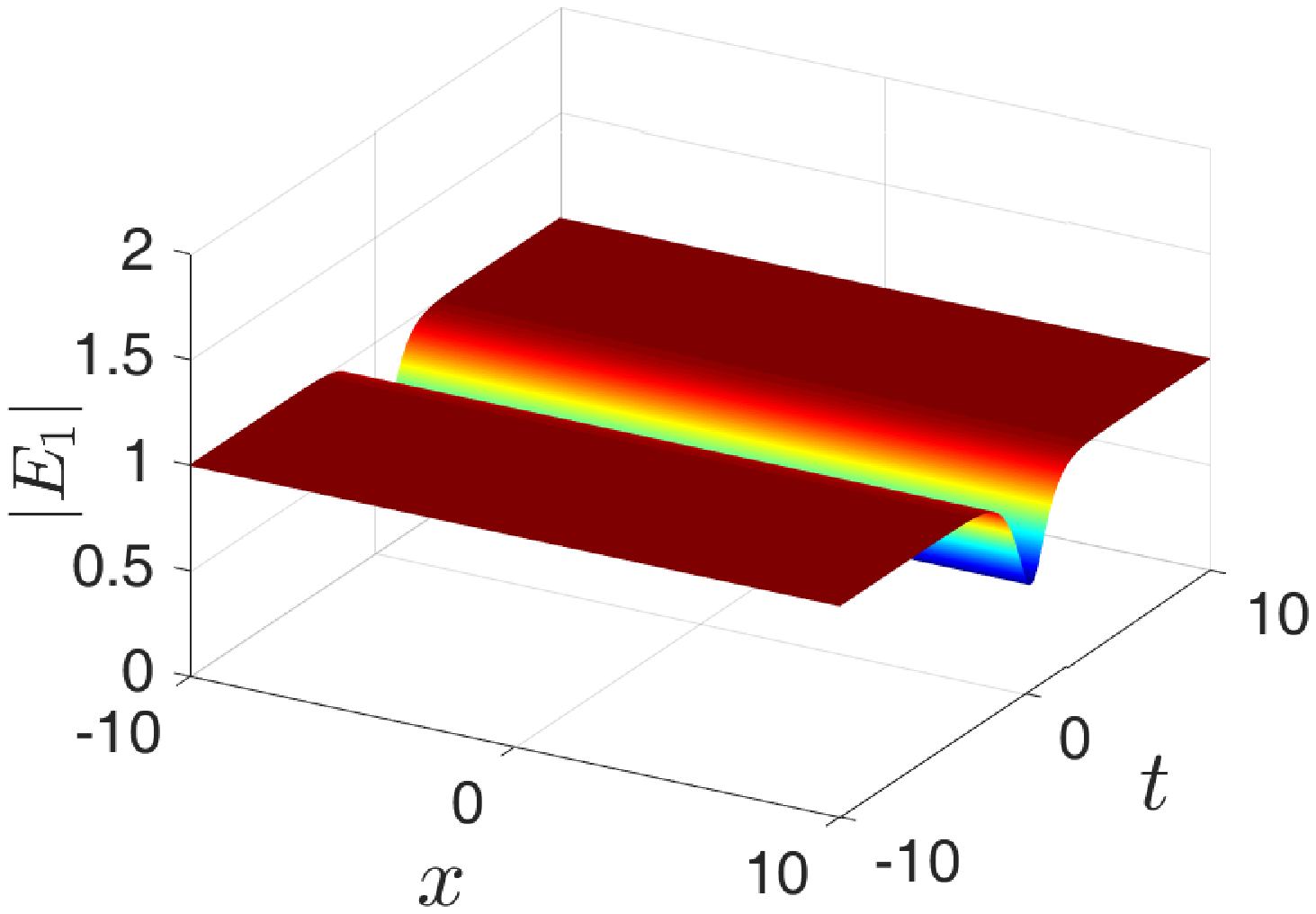}%
\includegraphics[height=100pt,width=80pt]{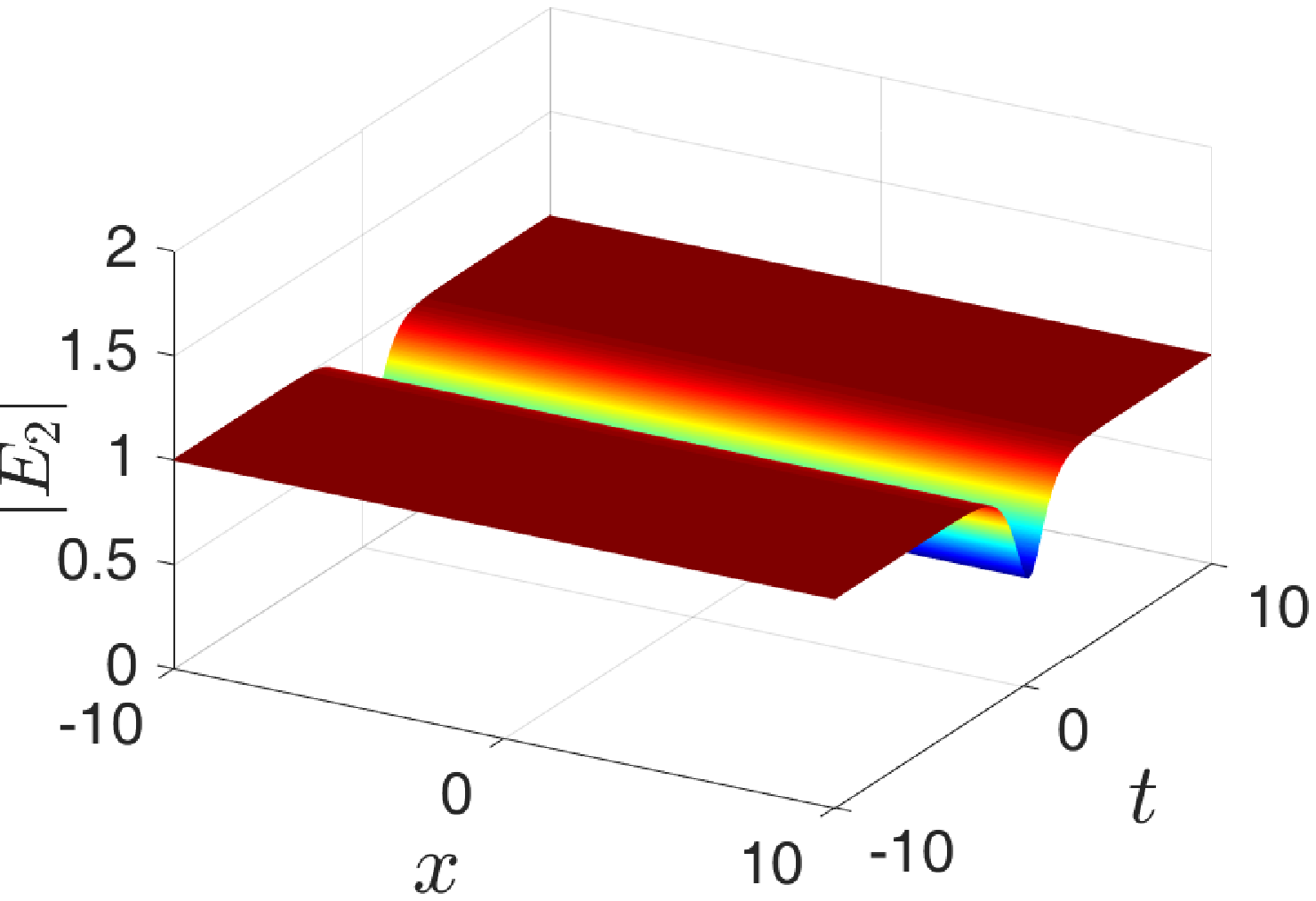} %
\includegraphics[height=100pt,width=80pt]{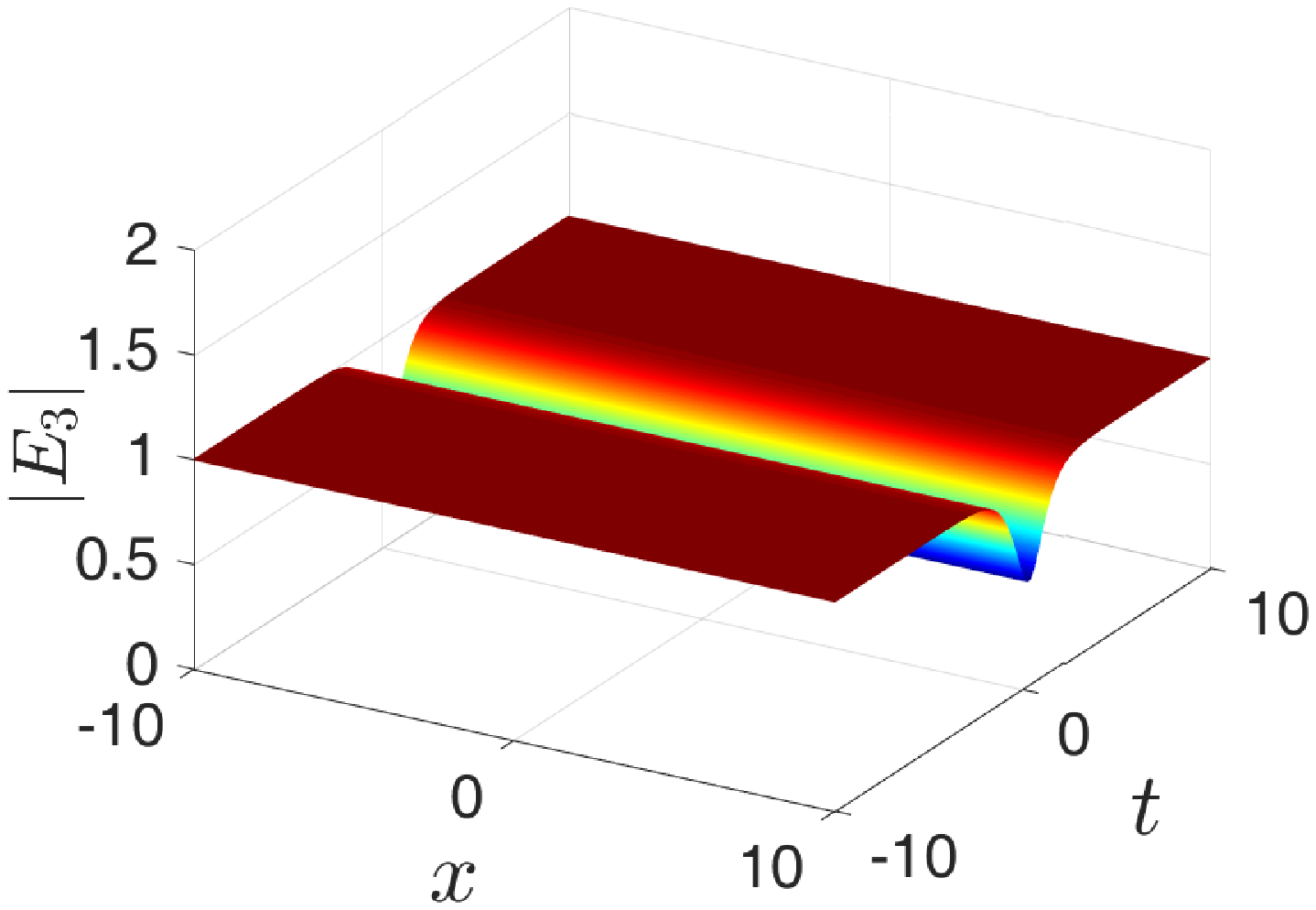}
\caption{An example of a time-localized dark mode produced by system~(%
\protect\ref{threeequ}) with parameters $\protect\sigma _{1}=\protect\sigma %
_{2}=\protect\sigma _{3}=1$, $V_{1}=2$, $V_{2}=1$, $a_{1}=a_{2}=a_{3}=1$, $%
\protect\eta _{1}^{(0)}=0$ and $p_{1}=\frac{1}{2}(\protect\sqrt{3}+i)$.}
\label{fig6}
\end{figure}

Following Ref.~\cite{LWB2022}, the condition of the ZWG MI for system~(\ref%
{threeequ}) is found as $(\gamma _{1}+\gamma _{2}-\gamma _{3})^{2}-4\gamma
_{1}\gamma _{2}<0$. Therefore, we conclude that the condition of the
occurrence of the ZWG MI is, once again, tantamount to the existence
condition for the time-localized dark mode. This solution is shown in Fig.~(%
\ref{fig6}).

\section{Conclusion}

The present work reveals the existence and origin of the unprecedented, to
our knowledge, species of dark and anti-dark quasi-soliton states, in the
form of the time-localized modes. Exact solutions of this type are produced
in two distinct integrable systems, \textit{viz}., the MTM (massive Thirring
model) and 3WRI (three-wave resonant-interaction) system, They provide
fundamental models for the propagation of nonlinear waves in media without
intrinsic dispersion, that find straightforward realizations in plasmas,
nonlinear optics, and hydrodynamics. In the MTM, the time-localized modes
feature a dark structure in one component and an anti-dark one in the other,
a feature that is explained on the basis of the associated norm-conservation
law. An important conclusion of the analysis is that the existence condition
for the time-localized modes in both models is tantamount to the condition
providing the occurrence of the ZWG MI (zero-wavenumber-gain modulational
instability). This is a natural conclusion, as it is the MI gain at the zero
modulation wavenumber, $Q=0$, that generates, respectively, the dip and
spike in the dark and anti-dark components of the mode. Our simulations
demonstrate that random perturbations, added to the CW background, give rise
to complex patterns composed of robust fragments in the form of the
time-localized modes. Furthermore, we have demonstrated that the
ZWG-MI-based mechanism creates the similar time-localized patterns (or
fractions thereof) in the non-integrable generalization of the MTM, which
includes the SPM terms, governing the light propagation in Bragg gratings.
Hence, it should be possible to create the predicted time-localized modes
experimentally in nonlinear optics. To illustrate the generality of the
predictions, an additional system featuring such time-localized modes was
also presented in the form of the three--wave resonant interaction system.

As a development of the present analysis, it will be relevant to study in
detail multi-soliton complexes of the time-localized type, as well as their
interactions 
with usual spatial solitons or rogue waves. The present study also suggests
that the search for time-localized modes in other ZWG-bearing systems is a
promising direction for future work. \vspace{2mm}

\section*{Acknowledgments}

This work has been supported by the National Natural Science Foundation of
China under Grant No.12205029 and by the Fundamental Research Funds of the
Central Universities (No. 230201606500048). The work of B.A.M. is supported,
in part, by the Israel Science Foundation (Grant No. 1695/22). The work of
P.G.K is supported by the US National Science Foundation under Grants No.
PHY-2110030 and DMS-2204702). 
\nocite{}

\end{document}